\documentclass[prd,twocolumn,showpacs,superscriptaddress]{revtex4}

\usepackage{amsfonts}
\usepackage{amsmath}
\usepackage{amssymb}
\usepackage{amsthm}
\usepackage{bm}
\usepackage{dcolumn}
\usepackage{epsfig}
\usepackage{graphicx}
\usepackage{graphics}
\usepackage[latin1]{inputenc}
\usepackage{latexsym}
\usepackage{rotating}
\usepackage{hyperref}

\newcommand\be{\begin{equation}}
\newcommand\ba{\begin{eqnarray}}
\newcommand\ee{\end{equation}}
\newcommand\ea{\end{eqnarray}}

\newcommand{\pont}{{\,^\ast\!}R\,R}

\begin{document}
\title{How do Black Holes Spin in Chern-Simons Modified Gravity?}

\author{Daniel Grumiller}
\affiliation{Center for Theoretical Physics,
  Massachusetts Institute of Technology, 77 Massachusetts Ave., Cambridge, MA 02139, USA}

\author{Nicol\'as Yunes}
\affiliation{Institute for Gravitation and the Cosmos, Department of Physics, 
The Pennsylvania State University, University Park, PA 16802, USA}

\date{\today}

\preprint{IGC-07/11-1}

%%%%%%%%%%%%%%%%%%%%%%%%%%%%%%%%%%%%%%%%%%%%%%%%%%%%%%%%%%%%%%%%%%%%%%%%%%%%%%
\begin{abstract}
  
  No Kerr-like exact solution has yet been found in Chern-Simons
  modified gravity. Intrigued by this absence, we study stationary and
  axisymmetric metrics that could represent the exterior field of
  spinning black holes.  For the standard choice of the background
  scalar, the modified field equations decouple into the Einstein
  equations and additional constraints.  These constraints eliminate
  essentially all solutions except for Schwarzschild. For
  non-canonical choices of the background scalar, we find several
  exact solutions of the modified field equations, including
  mathematical black holes and pp-waves. We show that the
  ultrarelativistically boosted Kerr metric can satisfy the modified
  field equations, and we argue that physical spinning black holes may
  exist in Chern-Simons modified gravity only if the metric breaks
  stationarity, axisymmetry or energy-momentum conservation.

\end{abstract}


%11.25.Wx       String and brane phenomenology
%04.80.Nn       Gravitational wave detectors and experiments (see also
%95.55.Ym-in astronomy) 
%04.60.-m       Quantum gravity
%04.80.Cc       Experimental tests of gravitational theories
%04.50.+h       Gravity in more than four dimensions, Kaluza-Klein
%theory, unified field theories; alternative theories of gravity
%(see also 11.25.Mj Compactification and four-dimensional models)
%11.25.Sq       Nonperturbative techniques; string field theory
%04.30.-w       Gravitational waves: theory
%04.65.+e       Supergravity (see also 12.60.Jv Supersymmetric models)
%04.30.Nk       Wave propagation and interactions

\maketitle

%%%%%%%%%%%%%%%%%%%%%%%%%%%%%%%%%%%%%%%%%%%%%%%%%%%%%%%%%%%%%%%%%%%%%%%%%%%%%%
\section{Introduction}
\label{intro}

General relativity (GR) is one of physics' most successful theories,
passing all experimental tests so far with ever increasing
accuracy~\cite{Will:2005va}.  Nevertheless, modifications to GR are
pursued vigorously for two main reasons: from a theoretical
standpoint, we search for an ultraviolet (UV) completion of GR, such
as string theory, that would lead to corrections in the action
proportional to higher powers of scalar invariants of the Riemann
tensor; from an experimental standpoint, observations in the deep
infrared (IR) regime suggest the existence of some form of dark
energy~\cite{Riess:1998cb,Perlmutter:1998np,Tegmark:2006az}. One
possibility to accommodate dark energy is to consider an action with
non-linear couplings to the Ricci
scalar~\cite{Jordan:1959eg,Brans:1961sx}, similar in spirit to the
corrections that we expect from a UV completion of GR.

UV and IR corrections entail higher derivatives of the fundamental
degrees of freedom in the equations of motion, which on general
grounds tend to have disastrous consequences on the stability of the
solutions of the theory~\footnote{If one considers the classical
  solutions of the non-modified theory (without higher derivatives)
  and regards additional terms as loop corrections, then no stability
  issues arise. In that case, additional (unstable) solutions must be
  considered spurious \cite{Simon:1990ic,Simon:1990jn}.}: the
so-called Ostrogradski instability (for a review
cf.~e.g.~\cite{Woodard:2006nt}). A few loopholes exist, however, that
allow to bypass this theorem (for example, if the non-linear
corrections can be converted into a representation of a scalar-tensor
theory).  Along these lines, special combinations of scalar invariants
that play the role of a topological term, such as the Euler or
Pontryagin term, can in general be added safely to the action.
 
In this paper, we study Chern-Simons (CS) modified
gravity~\cite{Jackiw:2003pm}, where the Einstein-Hilbert action is
modified by the addition of a parity-violating Pontryagin term. As
described by Jackiw and Pi~\cite{Jackiw:2003pm}, this correction
arises through the embedding of the $3$-dimensional CS topological
current into a $4$-dimensional spacetime manifold. CS gravity is not a
random extension of GR, but it has physical roots in particle physics.
Namely, if there is an imbalance between left- ($N_L$) and
right-handed ($N_R$) fermions, then the fermion number current $j^\mu$
has a well-known gravitational anomaly \cite{AlvarezGaume:1983ig},
$\partial_\mu j^\mu \propto (N_L-N_R) \pont$, analogous to the
original triangle anomaly \cite{Bell:1969ts}. Here $\pont$ is the
Pontryagin term (also known as the gravitational instanton density or
Chern-Pontryagin term) to be defined in the next section.  CS gravity
is also motivated by string theory: it emerges as an anomaly-canceling
term through the Green-Schwarz mechanism~\cite{Green:1987mn}.  Such a
correction to the action is indispensable, since it arises as a
requirement of all $4$-dimensional compactifications of string theory
in order to preserve unitarity~\cite{Alexander:2004xd}.

CS gravity has been studied in the context of cosmology, gravitational
waves, solar system tests and Lorentz invariance. In particular, this
framework has been used to explain the anisotropies in the cosmic
microwave background~\cite{Lue:1998mq,Li:2006ss,Alexander:2006mt} and
the leptogenesis problem~\cite{alexander:2004:lfg,Alexander:2004xd}
(essentially using the gravitational anomaly described above in the
other direction). CS gravity has also been shown to lead to amplitude
birefringent gravitational
waves~\cite{Jackiw:2003pm,Alexander:2004wk,Alexander:2007zg,Alexander:2007vt},
possibly allowing for a test of this theory with gravitational-wave
detectors~\cite{Alexander:2007:bgw}. Moreover, CS gravity has been
investigated in the far-field of a spinning binary system, leading to
a prediction of gyromagnetic
precession~\cite{Alexander:2007zg,Alexander:2007vt} that differs from
GR. This prediction was later improved on and led to a constraint on
the magnitude of the CS coupling~\cite{Smith:2007jm}. Finally, CS
gravity has been studied in the context of Lorentz-invariance and
-violation~\cite{Guarrera:2007tu} and the theory has been found to
preserve this symmetry, provided the CS coupling is treated as a
dynamical field. For further studies of these and related issues
cf.~e.g.~\cite{Jackiw:2003pm,Kostelecky:2003fs,Mariz:2004cv,Alexander:2004wk,Bluhm:2004ep,Eling:2004dk,Alexander:2004us,Lyth:2005jf,Mattingly:2005re,Lehnert:2006rp,Alexander:2006mt,Hariton:2006zj,Alexander:2007qe,Guarrera:2007tu,Alexander:2007vt,Konno:2007ze,Smith:2007jm,Fischler:2007tj,Tekin:2007rn}.
%and Refs.~therein.

CS gravity introduces the following modification to the
action~\cite{Jackiw:2003pm}: $S=S_{EH}+S_{\rm mat}+S_{CS}$, where
$S_{EH}$ is the Einstein-Hilbert action, $S_{\rm mat}$ is some matter
action, and the new term is given by
\begin{equation}
\label{eq:intro}
S_{CS} \sim \int dV  \;  \theta \; \pont \,.
\end{equation}
In Eq.~(\ref{eq:intro}), $dV$ is a $4$-dimensional volume element,
$\pont$ is the Pontryagin term and $\theta$ is a {\emph{background
    scalar field}} (we shall define this action in more detail in the
next section).  This scalar field, sometimes called a gravitational
axion, acts as a {\emph{CS coupling function}} that can be interpreted
either as an external or a dynamical quantity. In the former case, CS
gravity is an effective theory that derives from some other, more
fundamental gravity theory that physically defines the scalar field.
In the latter case, the scalar field possesses its own equation of
motion, which could in principle contain a potential and a kinetic
term~\cite{Smith:2007jm}.

The strength of the CS correction clearly depends on the CS coupling
function. If we consider CS gravity as an effective theory, the
coupling function is suppressed by some mass scale, which could lie
between the electro-weak and the Planck scale, but it is mostly
unconstrained~\cite{Smith:2007jm}. In the context of string theory,
the coupling constant has been computed in very conservative
scenarios, leading to a Planck mass suppression
\cite{alexander:2004:lfg}. In less conservative scenarios, there could
exist enhancements that elevate the coupling function to the realm of
the observable. Some of these scenarios are cosmologies where the
string coupling vanishes at late
times~\cite{Brandenberger:1988:sit,Tseytlin:1991:eos,Nayeri:2005:pas,sun:2006:ccm,wesley:2005:cct,Alexander:2000:bgi,Brandenberger:2001:lpi,battefeld:2006:sgc,brandenberger:2006:sgc,brandenberger:2007:sgc,brax:2004:bwc},
or where the field that generates $\theta$ couples to spacetime
regions with large curvature~\cite{randall:1999:lmh,randall:1999:atc}
or stress-energy density~\cite{Alexander:2007:bgw}.

The CS correction is encoded in the modified field equations, which
can be obtained by varying the modified action with respect to the
metric. The divergence of the modified field equations establishes the
Pontryagin constraint $\pont=0$, through the Bianchi identities for a
vacuum or conserved stress-energy tensor. Not only does this
constraint have important consequences on the conservation of energy,
but it also restricts the space of solutions of the modified theory.
For example, although this restriction is not strong enough to
eliminate the Schwarzschild solution, it does eliminate the Kerr
solution. Since astrophysical observations suggest that supermassive
black holes (BHs) at the center of galaxies do have a substantial spin
(cf.,~e.~g.~\cite{Aschenbach:2007cr} and references therein), this
raises the interesting question of what replaces the Kerr solution in
CS gravity.

%%%%%%%%%%%%%%%%%%%%%%%%%%%%%%%%%%%%%%%%%%%%%%%%%%%%%%%%%%%%%%%%%%%%
% RESULTS
%%%%%%%%%%%%%%%%%%%%%%%%%%%%%%%%%%%%%%%%%%%%%%%%%%%%%%%%%%%%%%%%%%%

In this paper, we search for solutions to the CS modified field
equations that could represent the exterior gravitational field of a
spinning star or BH.  We find that solutions cluster into two
different classes: GR solutions that independently satisfy both the
vacuum Einstein equations and the modified field equations; non-GR
solutions that satisfy the modified field equations but not the vacuum
Einstein equations. We carry out an extensive study of solutions by
looking at three groups of line elements: spherically symmetric
metrics; static and axisymmetric metrics; and stationary and
axisymmetric metrics. The first group contains GR solutions only,
independently of the choice of the CS scalar field.  The second group
leads to a decoupling of the modified field equations for 'natural'
choices of the scalar field, which again reduces to trivial GR
solutions.  In fact, we show here that static and axisymmetric line
elements are forced to be spatially conformally flat if such a
decoupling occurs. The third group also leads to the same decoupling
for the canonical choice of the scalar field, and we argue against the
existence of non-trivial solutions.

This paper suggests that stationary and axisymmetric line elements in
CS gravity probably do not admit solutions of the field equations for
the canonical choice of the CS scalar field. However, solutions do
exist when more general scalar fields are considered, albeit not
representing physical BH configurations~\footnote{The conclusion that
  stationary and axisymmetric solutions to the modified field
  equations do not exist is in agreement with~\cite{Konno:2007ze}.
  However, this fact does not lead to the conclusion that spinning BHs
  cannot exist in the modified theory, as implied
  in~\cite{Konno:2007ze}. In fact, an approximate solution (albeit not
  axisymmetric) that can represent a spinning BH in CS gravity has
  already been found in the
  far-field~\cite{Alexander:2007vt,Smith:2007jm}.}.  We find two types
of solutions, mathematical BHs and ultrarelativistically boosted BHs,
which, to our knowledge, are the first examples of BH and BH-like
solutions in CS gravity, besides Schwarzschild and
Reissner-Nordstr\"om. The first type arises when we consider a
subclass of stationary and axisymmetric line elements (the so-called
van Stockum class), for which we find both GR and non-GR solutions for
non-canonical scalar fields. For instance, we shall demonstrate that
the line-element
\be
\label{eq:conclusion1}
ds^2 = -\rho \Big(1-\frac{2m}{\sqrt{\rho}}\Big) dt^2 - 2 \rho \,dt\, d\phi + \frac{1}{\sqrt{\rho}} \Big(d\rho^2 + dz^2\Big)\,,
\ee
together with the CS scalar field $\theta = 2\sqrt{\rho}\,z/3$,
satisfies the modified field equations but does not arise in GR as a
vacuum solution.  The metric in Eq.~\eqref{eq:conclusion1} represents
BHs in the mathematical sense only: it exhibits a Killing horizon at
$\sqrt{\rho}=2m=\rm const.$, but it contains unphysical features, such
as (naked) closed time-like curves.  The second type of solutions with
a non-canonical scalar field arises when we consider scalar fields
whose divergence is a Killing vector. These fields lead to exact
gravitational pp-wave solutions of GR and non-GR type. One particular
example that we shall discuss in in this paper is the
ultrarelativistically boosted Kerr BH,
\be
ds^2=-2du\,dv - h_0 \delta(u) \ln\left(x^2 + y^2\right)du^2+dx^2+dy^2\,,
\label{eq:intro17}
\ee
with the CS scalar field $\theta = \lambda v$, where $h_0$ and
$\lambda$ are constants. 

Although we did not find a Kerr analogue by searching for stationary
and axisymmetric solutions, spinning BHs do seem to exist in the
theory. This suggestion is fueled by the existence of two different
limits of the Kerr spacetime that are still preserved: the
Schwarzschild limit and the Aichelburg-Sexl limit,
Eq.~\eqref{eq:intro17}, which we shall show persists in CS gravity.
These limits, together with the existence of a non-axisymmetric
far-field solution~\cite{Alexander:2007vt}, indicate that a spinning
BH solution must exist, albeit not with the standard symmetries of the
Kerr spacetime. Unfortunately, spacetimes with only one or no Killing
vector are prohibitively general and their study goes beyond the scope
of this work. Nonetheless, the possibility of constructing such
solutions by breaking stationarity or axisymmetry is discussed and a
better understanding of solutions in CS gravity is developed. Finally,
we show how to recover the Kerr solution by postulating, in an ad-hoc
manner, a non-conserved energy momentum-tensor and deduce that it
violates the classical energy conditions.

%%%%%%%%%%%%%%%%%%%%%%%%%%%%%%%%%%%%%%%%%%%%%%%%%%%%%%%%%%%%%%%%%%%%%%%5

This paper is organized as follows: Sec.~\ref{ABC} reviews some basic
features of CS modified gravity and exploits two alternative
formulations of the Pontryagin constraint, one based upon the
spinorial decomposition of the Weyl tensor and one based upon its
electro-magnetic decomposition, to reveal some physical consequences of this constraint; Sec.~\ref{persistence} revisits the
Schwarzschild, Friedmann-Robertson-Walker and
Reissner-Nordstr\"om solutions in CS modified gravity and
addresses the sensitivity of these solutions to the choice of CS
coupling function; Sec.~\ref{static} studies static, axisymmetric line
elements in great detail, while Sec.~\ref{stationary} investigates
stationary, axisymmetric metrics and provides the first non-trivial
exact solutions to CS modified gravity, including mathematical BH
solutions; Sec.~\ref{beyond} addresses metrics that break axisymmetry
or stationarity and concentrates on non-trivial solutions for pp-waves
and the Aichelburg-Sexl boosted Kerr metric; Sec.~\ref{conclusions}
concludes and points to future research.

We use the following conventions in this paper: we work exclusively in
four spacetime dimensions with signature
$(-,+,+,+)$~\cite{Misner:1973cw}, with Latin letters $(a,b,\ldots,h)$
ranging over all spacetime indices; curvature quantities are defined
as given in the MAPLE GRTensorII package~\cite{grtensor}; round and
square brackets around indices denote symmetrization and
anti-symmetrization respectively, namely $T_{(ab)}:=\frac12
(T_{ab}+T_{ba})$ and $T_{[ab]}:=\frac12 (T_{ab}-T_{ba})$; partial
derivatives are sometimes denoted by commas, e.g.~$\partial
\theta/\partial r=\partial_r\theta=\theta_{,r}$. The Einstein
summation convention is employed unless otherwise specified, and we
use geometrized units where $G=c=1$.

%%%%%%%%%%%%%%%%%%%%%%%%%%%%%%%%%%%%%%%%%%%%%%%%%%%%%%%%%%%%%%%%%%%%%%%%%%%
\section{CS modified gravity}
\label{ABC}

%-------------------------------------------------------------------
\subsection{ABC of CS}\label{se:ABC}
  
In this section, we summarize the basics of CS modified gravity,
following the formulation of~\cite{Jackiw:2003pm}. Let us begin by
defining the full action of the theory~\footnote{There is a relative
  sign difference in the CS correction to the action compared
  to~\cite{Jackiw:2003pm}. This minus sign is included in order to
  obtain the same equations of motion as in~\cite{Jackiw:2003pm},
  correcting a minor typo.}:
\be
\label{CSaction}
S = \kappa \int d^4x \sqrt{-g} \left(R - \frac{1}{4} \theta \; \pont
\right) +S_{\rm mat}\,, 
\ee
where $\kappa = 1/(16 \pi)$, $g$ is the determinant of the metric, the
integral extends over all spacetime, $R$ is the Ricci scalar, $S_{\rm
  mat}$ is some unspecified matter action and $\pont$ is the
Pontryagin term. The latter is defined via
\be
\label{pontryagindef}
\pont:={\,^\ast\!}R^a{}_b{}^{cd} R^b{}_{acd}\,,
\ee
where the dual Riemann-tensor is given by
\be
\label{Rdual}
{^\ast}R^a{}_b{}^{cd}:=\frac12 \epsilon^{cdef}R^a{}_{bef}\,,
\ee
with $\epsilon^{cdef}$ the 4-dimensional Levi-Civita
tensor~\footnote{We prefer to work with tensors rather than with
  tensor densities in this paper, so some expressions might appear to
  differ by factors of $\sqrt{-g}$ from \cite{Jackiw:2003pm}.}. The
Pontryagin term [Eq.~\eqref{pontryagindef}] can be expressed as the
divergence
\be
\nabla_a K^a = \frac14 \pont 
\label{eq:curr1}
\ee
of the Chern-Simons topological current ($\Gamma$ is the Christoffel
connection), 
\be
K^a :=\epsilon^{abcd}\left(\Gamma^n{}_{bm}\partial_c\Gamma^m{}_{dn}+\frac23\Gamma^n{}_{bm}\Gamma^m{}_{cl}\Gamma^l{}_{dn}\right)\,,
\label{eq:curr2}   
\ee
thus the name ``Chern-Simons modified gravity''~\footnote{If $\nabla_a
  K^a$ is converted into $1/\sqrt{g} \, \partial_a (\sqrt{g} K^a)$ the
  results (2.4) and (2.5) of \cite{Jackiw:2003pm} are recovered.}.

The modified field equations can be obtained by varying the action
with respect to the metric. Exploiting the well-known relations
\be
\delta R^b{}_{acd}=\nabla_c\delta\Gamma^b{}_{ad}-\nabla_d\delta\Gamma^b{}_{ac}
\ee
and
\be
\delta\Gamma^b{}_{ac} = \frac12 g^{bd}\left(\nabla_a\delta
  g_{dc}+\nabla_c\delta g_{ad}-\nabla_d\delta g_{ac}\right)\,, 
\ee
the variation of the geometric part of the action leads to
\ba
\label{variationofS}
\delta S -\delta S_{\rm mat} &=& \kappa \int d^4x \sqrt{-g}
  \left(R_{a b} - \frac{1}{2} g_{ab} R + C_{ab} \right) \delta g^{ab} 
\nonumber \\
&-& \frac{\kappa}{4} \int d^4x \sqrt{-g}  \pont \; \delta\theta
\nonumber \\
&+& \Sigma_{EH} + \Sigma_{CS}\,.
\ea
Here, the tensor $C_{ab}$ stands for a
$4$-dimensional Cotton-like tensor, which we shall refer to as the
C-tensor~\footnote{In the original work of~\cite{Jackiw:2003pm}, this
  tensor was called 'Cotton tensor' because it shares
  similarities with the $3$-dimensional Cotton-tensor. However, the
  notion of a higher-dimensional Cotton tensor already
  exists~\cite{Garcia:2003bw} and differs from the definition of
  $C_{ab}$, which is why we refer to Eq.~(\ref{Ctensor}) as a
  ``C-tensor''.}, given by
\be
\label{Ctensor}
C^{ab} := v_c
\epsilon^{cde(a}\nabla_eR^{b)}{}_d+v_{cd}{\,^\ast\!}R^{d(ab)c}\,,   
\ee
where 
\be
\label{v}
v_a:=\nabla_a\theta\,,\qquad
v_{ab}:=\nabla_a\nabla_b\theta=\nabla_{(a}\nabla_{b)}\theta 
\ee 
are the velocity and covariant acceleration of $\theta$,
respectively~\footnote{The quantity $v_a$ is sometimes referred to as
  an {\emph{embedding coordinate}} since it embeds the $3$-dimensional
  CS theory into a $4$-dimensional spacetime.}.  We shall always
assume that $v_a$ does not vanish identically, because otherwise the
model reduces to GR~\footnote{When $v_{a} = 0$ then $\theta$ is constant
  and the Pontryagin term becomes a topological term not contributing to the
  field equations.}.

Surface terms are collected in the third line of
Eq.~(\ref{variationofS}) and arise due to repeated integration by
parts and application of Stokes' theorem. In particular, $\Sigma_{EH}$
and $\Sigma_{CS}$ arise from variation of the Einstein-Hilbert and CS
sector of the action, respectively. The former expression is
well-known, while the latter contains a term with $\delta\Gamma$,
\be
\label{eq:boundary}
\Sigma_{CS} = \kappa\int d^4x\sqrt{-g} \nabla_d\left(\theta {\,^\ast\!}R^{abcd}\delta\Gamma_{bac}\right) + \dots
\ee
It is worthwhile pointing out
that one cannot just impose Dirichlet boundary conditions on the
induced metric at the boundary by adding the Gibbons-Hawking-York
term, as it is the case in GR \cite{York:1972sj,Gibbons:1976ue}. There is no obvious way to cancel the
term containing the variation of the connection, $\delta\Gamma$, in
Eq.~\eqref{eq:boundary}, except by imposing suitable fall-off
conditions on the scalar field $\theta$ or Dirichlet boundary
conditions on the connection. Even though we shall neglect boundary
issues henceforth, we emphasize that these considerations are relevant
in many applications, such as BH thermodynamics. 

The modified field equations are then given by the first line of
Eq.~(\ref{variationofS}), provided the second line vanishes. The
vanishing of $\pont$ is the so-called Pontryagin constraint and we
shall study it in Sec.~\ref{se:pont}. The modified field equations in
the presence of matter sources are then given by
\be
 G_{ab} + C_{ab} = 8 \pi T_{ab},
\label{eq:eom}
\ee
where $G_{ab}=R_{ab}-\frac12 g_{ab}R$ is the Einstein tensor and $T_{ab}$ is the
stress-energy tensor of the source. In this paper, we are primarily
concerned with the vacuum case, $T_{ab}=0$, for which the
modified field equations reduce to
\be
\label{eom}
R_{ab} + C_{ab} = 0,
\ee
due to the tracelessness of the C-tensor, $C^a{}_a=0$. Like in GR, vacuum solutions in CS gravity satisfy 
\be
R = 0\,.
\ee

%-----------------------------------------------------------------
\subsection{Pontryagin Constraint}\label{se:pont}

Let us now discuss the Pontryagin constraint
\be 
\label{eq:constraint}
\pont = 0,
\ee
which then forces the second line in Eq.~\eqref{variationofS} to
vanish. One route to obtain the Pontryagin constraint is to treat
$\theta$ as a dynamical field (or rather a Lagrange multiplier). By
varying the action with respect to $\theta$, we obtain the equations
of motion for the scalar field that dynamically enforce the Pontryagin
constraint.

Another route to obtain the Pontryagin constraint is to treat $\theta$
as an external quantity. In this case, there are no equations of
motion for the scalar field. Nonetheless, by taking the covariant
divergence of the equations of motion and using the contracted Bianchi
identities, one obtains
\be
\label{nablaC}
\nabla_a C^{ab}=\frac18 v^b \pont = 8\pi \nabla_a T^{ab}.
\ee
Usually, it is desirable to require that the stress-energy be
covariantly conserved. However, in CS modified gravity this need not
be the case because a non-vanishing covariant divergence $\nabla_a
T^{ab}\neq 0$ could be balanced by a non-vanishing Pontryagin term --
this is, in fact, how the term arises in some approaches in the first
place, cf.~\cite{Alexander:2004us}. We shall come back to this issue
at the end of Sec.~\ref{beyond}, but for the time being we shall set
$T^{ab}=0$, which then leads to the Pontryagin constraint.

The Pontryagin constraint is a necessary condition for any vacuum
spacetime that solves the modified field equations, but what does it
mean physically? We shall attempt to answer this question by providing
two alternative formulations of this constraint, but before doing so,
let us discuss some general properties and consequences of
Eq.~\eqref{eq:constraint}. First, notice that setting the $\pont$ term
to zero leads to the conserved current $K^{a}$ [Eqs.~\eqref{eq:curr1}
and~\eqref{eq:curr2}], which is topological in nature, and thus
implies this quantity is intrinsically different from typical
conserved quantities, such as energy or angular momentum.  Second,
when the CS action is studied on-shell [Eq.~(\ref{CSaction}) with
$\pont = 0$] it reduces to the GR action, an issue that is of
relevance for stability considerations, e.g.~thermodynamic stability
in BH mechanics.

The first physical interpretation of the Pontryagin constraint can be
obtained by considering a spinorial decomposition.  Let us then
consider the useful relation
\be
\pont=\,{^\ast\!}C\,C\,,
\label{eq:ha!}
\ee
which we prove in appendix \ref{app:new}. In Eq.~(\ref{eq:ha!}), $C$
is the Weyl tensor defined in \eqref{eq:appnew2} and $\,{^\ast\!}C$
its dual, defined in \eqref{Cdual}. This identity allows us to use 
powerful spinorial methods to map the Weyl tensor into the Weyl spinor
\cite{Penrose:1986ca}, which in turn can be characterized by the
Newman-Penrose (NP) scalars
$\left(\Psi_0,\Psi_1,\Psi_2,\Psi_3,\Psi_4\right)$. In the notation of
\cite{Stephani:2003tm}, the Pontryagin constraint translates into a
reality condition on a quadratic invariant of the Weyl spinor,
${\cal{I}}$,
\be
\label{I2}
\Im \left({\cal{I}}\right) = \Im{\left(\Psi_0\Psi_4+3\Psi_2^2-3\Psi_1\Psi_3\right)}=0\,.
\ee 

Such a reality condition is particularly useful for the consideration
of algebraically special spacetimes. For instance, it follows
immediately from Eq.~(\ref{I2}) that spacetimes of Petrov types $III$,
$N$ and $O$ obey the Pontryagin constraint, since in the latter case
all NP scalars vanish, while in the former cases (in an adapted frame)
only $\Psi_3$ or $\Psi_4$ are non-vanishing. Moreover, all spacetimes
of Petrov types $D$, $II$ and $I$ are capable of violating
Eq.~\eqref{I2}. For example, for spacetimes of Petrov type $II$ one
can choose an adapted tetrad such that $\Psi_0=\Psi_1=\Psi_3=0$, which
then reduces Eq.~\eqref{I2} to the condition that either the real part
or the imaginary part of $\Psi_2$ has to vanish. 

The reality condition of Eq.~\eqref{I2} can also be useful in
applications of BH perturbation theory. For instance, in the metric
reconstruction of the perturbed Kerr spacetime~\cite{Yunes:2005ve},
the NP scalars $\Psi_1=\Psi_3=0$ vanish.  In this context
gravitational waves are characterized by $\Psi_{4,0}$, while $\Psi_2$
is in general non-vanishing. In a tetrad that represents a
transverse-traceless frame, these scalars are given by
\be
\Psi_{4,0} = \ddot{h}_+ \mp i \ddot{h}_{\times},
\ee
where $h_{+,\times}$ are the plus/cross polarization of the waveform,
and the overhead dot stands for partial time
derivative~\cite{Buonanno:2006ui}. Obviously, $\Psi_0 \Psi_4 =
(\ddot{h_{+}})^2 + (\ddot{h_{\times}})^2$ is real, which again reduces
Eq.~\eqref{I2} to the condition that either the real part or the
imaginary part of $\Psi_2$ has to vanish. Neither of these
possibilities is the case for the Kerr BH or perturbations of
it~\cite{Yunes:2005ve}.
 
Another interpretation of the Pontryagin constraint can be obtained by
exploiting the split of the Weyl tensor into electric and magnetic
parts (cf.~e.g.~\cite{Cherubini:2003nj}).  Given some time-like vector
field $u^a$, normalized so that $u_au^a=-1$, one can define the
electric and magnetic parts of the Weyl tensor as
\be
\label{elecmagn}
(C_{abcd}+\frac i2 \epsilon_{abef}C^{ef}{}_{cd})u^bu^d=E_{ac}+iB_{ac}\,.
\ee
Then, the Pontryagin constraint is equivalent to the
condition~\footnote{One can show that Eq.~\eqref{EB} is related to the
  vanishing of certain derivatives of the Regge-Wheeler function in
  the Regge-Wheeler~\cite{Regge:1957rw} decomposition of the metric
  perturbation. We are currently studying how such a condition impacts
  the ringing of CS BHs elsewhere~\cite{nico:carlos}. The equivalence
  between Eqs.~\eqref{I2} and \eqref{EB} was shown for the first time
  in~\cite{Matte:1953}.}
\be
\label{EB}
E_{ab} B^{ab}=0\,.
\ee
This leads to three possibilities: either the spacetime is purely
electric ($B_{ab}=0$) or purely magnetic ($E_{ab}=0$) or orthogonal,
in the sense that Eq.~\eqref{EB} holds. Equation~\eqref{EB} is a
perfect analogue to the condition ${^\ast\!}F\,F\propto\bm{E \cdot
  B}=0$, which holds for specific configurations in electro-dynamics,
including purely electric ($\bm{B}=0$), purely magnetic ($\bm{E}=0$)
and electromagnetic wave configurations ($\bm{E}\neq 0 \neq \bm{B}$,
$\bm{E \cdot B}=0$). This suggests that there could be single
shock-wave solutions in CS gravity compatible with Eq.~\eqref{EB},
which we shall indeed encounter in Sec.~\ref{beyond}. In light of this
electro-magnetic analogy, the Pontryagin constraint can be rephrased
as ``the gravitational instanton density must vanish,'' since the
quantity ${^\ast\!}F\,F$ is sometimes referred to as the ``instanton
density.''

The electromagnetic decomposition of the Pontryagin constraint also
allows for a physical interpretation in terms of perturbations of the
Schwarzschild solution. In BH perturbation theory
(cf.~e.g.~\cite{Poisson}), the metric perturbation is also decomposed
through the electromagnetic Weyl tensor. The electric and magnetic
parts can then be related to the flux of mass and angular momentum
across the horizon.  Suffice it to say that for a binary BH system in
the slow-motion/small-hole approximation~\cite{Poisson}, these tensors
are of order $E_{ab} \sim {\cal{O}}(\Phi)$ and $B_{ab} \sim
{\cal{O}}(v \,\Phi)$, where the Newtonian potential $\Phi$ is of
${\cal{O}}(v^2)$ via the Virial theorem, with $v \ll 1$ the orbital
velocity. In this case, the Pontryagin constraint is satisfied
automatically up to terms of ${\cal{O}}(v^5)$. Within the
post-Newtonian (PN) approximation~\cite{Blanchet:2002av}, these
conclusions imply that the PN metric for non-spinning point-particles
in the quasi-circular approximation violates the Pontryagin constraint
at $2.5$ PN order [${\cal{O}}(v/c)^5$], which is precisely the order
at which gravitational waves appear.

Even for non-canonical choices of the scalar field, such as $\theta =
z$ proposed in~\cite{Konno:2007ze}, the far field expansion of the
Kerr metric does not satisfy the Pontryagin constraint to all orders.
This is so because obviously $\pont$ is independent of $\theta$. In
fact, one can show that violations of the constraint for the metric
considered in~\cite{Konno:2007ze} occur already at second order in the
metric perturbation, which renders this metric hopeless as an exact CS
solution. This observation is concurrent with the role the Pontryagin
constraint may play for gravitational waves \cite{Lehnert:2007}.

Finally, we can employ the electromagnetic analogy to anticipate the
answer to the question we pose in the title of this paper. Namely, we
are looking for a ``rotating charge'' configuration (where $\bm{E}\neq
0\neq \bm{B}$), which simultaneously is an ``electromagnetic wave''
configuration (where $\bm{E \cdot B}=0$). We know that no such
solutions exist in electrodynamics, except for two limits
\footnote{Actually there is a third limit, $\bm{E}=0$, which is either
  trivial (if also $\bm{B}$ vanishes) or a magnetic monopole.}: if the
rotation (and thus $\bm{B}$) approaches zero or if the charge is
infinitely boosted (and thus $\bm{B}$ becomes orthogonal to $\bm{E}$).
The first case corresponds to a static configuration, while the second
one to an ultrarelativistic limit. We shall indeed find below both
analogues as solutions of CS modified gravity, but we stress that the
naive analogy with electrodynamics does not yet rule out other
possible spinning configurations in CS modified gravity.

%------------------------------------------------------------------
\subsection{Space of Solutions}\label{se:sol}

\begin{figure}
\begin{center}
\includegraphics[scale=0.4,clip=true]{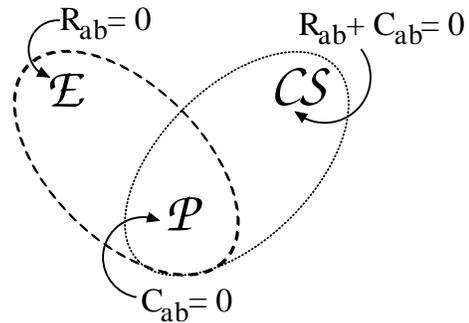}
\caption{\label{sol-space} 
  Space of solutions of Einstein gravity ${\cal{E}}$ and CS modified
  gravity ${\cal{CS}}$.}
\end{center}
\end{figure}

Before discussing some specific solutions to the CS modified field
equations, let us classify the space of solutions.
Figure~\ref{sol-space} presents a 2-dimensional depiction of this
space. The set ${\cal{E}}$ denotes the Einstein space of solutions,
whose members have a vanishing Ricci tensor, while ${\cal{CS}}$
denotes the CS space of solutions, whose members satisfy the CS
modified field equations [Eq.~\eqref{eom}], without necessarily being
Ricci flat. The intersection of the Einstein and the CS space defines
the Pontryagin space, denoted by ${\cal{P}} := {\cal{E}} \cap
{\cal{CS}}$, whose members satisfy both the Einstein and the modified
field equations independently. Therefore, solutions that live in
${\cal{P}}$ possess a vanishing C-tensor and automatically satisfy the
Pontryagin constraint, while those living in ${\cal{E}} \; \backslash
\; {\cal{P}}$ satisfy the vaccum Einstein equations but not the
Pontryagin constraint. Moreover, solutions that live in ${\cal{CS}} \;
\backslash \; {\cal{P}}$ are not Ricci-flat but do satisfy the
Pontryagin constraint because they solve the modified field equations.
Solutions of class ${\cal{P}}$ shall be referred to as GR solutions,
while solutions of class ${\cal{CS}} \; \backslash \; {\cal{P}}$ shall
be referred to as non-GR solutions.

To date, only one non-GR solution has been found perturbatively
\cite{Alexander:2007vt} by assuming a far field expansion for
point-particle sources in the PN weak-field/slow-motion approximation.
We shall show in the next sections that non-GR solutions exist only in
scenarios with a sufficient degree of generality, but not in highly
symmetric cases.  In the language of dynamical systems theory, the
${\cal{P}}$ space acts as an ``attractor'' of highly symmetric
solutions, emptying out the ${\cal{CS}}$ space.

In view of this, let us discuss some properties of solutions that live
in the ${\cal{P}}$ space. In this space, the C-tensor simplifies to
\be
\label{Cvacuum}
C^{ab}|_{\rm R_{ab}=0} = v_{cd}{\,^\ast\!}R^{d(ab)c} =
v_{cd}{\,^\ast\!}C^{d(ab)c}=0\,, 
\ee
where $C_{abcd}$ is the Weyl tensor and ${\,^\ast\!}C$ its dual,
defined in Eqs.~\eqref{eq:appnew2} and~\eqref{Cdual}.
Equation~\eqref{Cvacuum} leads to three distinct possibilities:
\begin{enumerate}
\item The (dual) Weyl tensor vanishes. However, since class
  ${\cal{P}}$ members also have a vanishing Ricci tensor, this
  condition reduces all possible solutions to Minkowski space.
\item The covariant acceleration of $\theta$ vanishes. This condition
  imposes a strong restriction on the geometry
  (cf.~e.g.~\cite{Stephani:2003tm}), which leads to spacetimes that
  are either flat or exhibit a null Killing vector.
\item Only the contraction of the covariant acceleration with the dual
  Weyl tensor vanishes.
\end{enumerate}
Moreover, for solutions in ${\cal{P}}$, the vanishing of the Ricci
tensor forces the Weyl tensor to be divergenceless, via the contracted
Bianchi identities. These observations are a clear indication that the
solutions inhabiting ${\cal{P}}$ must be special -- for instance,
exhibit a certain number of Killing vectors. Conversely, one may
expect that solutions inhabiting ${\cal{CS}} \; \backslash \;
{\cal{P}}$ cannot be ``too special.'' We shall put these expectations
on a solid basis and confirm them in the next sections.

%%%%%%%%%%%%%%%%%%%%%%%%%%%%%%%%%%%%%%%%%%%%%%%%%%%%%%%%%%%%%%%%%%%%
\section{Persistence of GR solutions}
\label{persistence}
%%%%%%%%%%%%%%%%%%%%%%%%%%%%%%%%%%%%%%%%%%%%%%%%%%%%%%%%%%%%%%%%%%%%
In this section, we study some solutions of GR that are known to
persist in CS gravity~\cite{Jackiw:2003pm,Guarrera:2007tu}, using the
insight on the Pontryagin constraint gained so far.  In the language
of Sec.~\ref{se:sol} we look for solutions that inhabit ${\cal{P}}$,
cf.~Fig.~\ref{sol-space}.

%-----------------------------------------------------------------
\subsection{Schwarzschild Solution}

The Schwarzschild solution, 
\begin{multline}
\label{Schw}
ds^2=-\left(1-\frac{2M}{r}\right)dt^2+\left(1-\frac{2M}{r}\right)^{-1}dr^2\\
+r^2\left(d\Theta^2+\sin^2{(\Theta)}d\phi^2\right)\,,
\end{multline}
is also a solution of the CS modified field equations if \cite{Jackiw:2003pm}
\be
\label{canon}
\theta = \frac{t}{\mu} \quad \rightarrow \quad v_{\mu} = [1/\mu,0,0,0]\,.
\ee
We refer to Eq.~\eqref{canon} as the canonical choice of the CS scalar
field~\cite{Jackiw:2003pm}.  In that case, the C-tensor can be
interpreted as a $4$-dimensional generalization of the ordinary
$3$-dimensional Cotton tensor.  Moreover, spacetime-dependent
reparameterization of the spatial variables and time translation
remain symmetries of the modified action~\cite{Jackiw:2003pm}.

We investigate now the most general
form of $\theta=\theta(t,r,\Theta,\phi)$ that will leave the
Schwarzschild metric a solution of the modified theory. The Pontryagin
constraint always holds, regardless of $\theta$, because the spacetime
is spherically symmetric, but $C_{ab}=0$ yields non-trivial equations.
Since we have chosen the Schwarzschild line element, we cannot force
the (dual) Weyl tensor to vanish (option $1$ in Sec.~\ref{se:sol}), where
the only linearly independent component is
\be
C_{t r \Theta \phi} = \frac{2 M}{r} \sin{\Theta}\,.
\ee
Another possibility is to force the scalar field to have a vanishing
covariant acceleration (option $2$ in Sec.~\ref{se:sol}). This condition
then yields 
%the following differential equations
%
%\ba
%0 &=& \theta_{,tt} r^3 - m r \theta_{,r} f\,,
%\nonumber \\
%0 &=& \theta_{,tr} r^2 f - \theta_{,t} m\,,
%\nonumber \\
%0 &=& \theta_{,r r} r^2 f + \theta_{,r} m\,,
%\nonumber \\
%0 &=& \theta_{,t\Theta}= \theta_{,t\phi}\,,
%\nonumber \\
%0 &=& \partial_r \left(\frac{\theta_{,\Theta}}{r}\right)=\partial_r \left(\frac{\theta_{,\phi}}{r}\right)\,,
%\nonumber \\
%0 &=& \theta_{,\Theta \Theta} + \theta_{,r} r f\,,
%\nonumber \\
%0 &=& \partial_{\Theta} \left(\frac{\theta_{,\phi}}{\sin{\Theta}}\right)\,,
%\nonumber \\
%0 &=& \theta_{,\phi \phi} + \theta_{,r} r \sin{\Theta}^2 f  +
%\theta_{,\Theta} \sin{\Theta} \cos{\Theta}\,,\nonumber
%\ea
%
%where $f = 1 - 2 m/r$ is the Schwarzschild factor. These equations form 
an over-constrained system of partial differential equations (PDEs), whose only solution for $M\neq 0$ is the
trivial one: constant $\theta$. 
We are thus left with the remaining possibility (option $3$ in
Sec.~\ref{se:sol}), namely that only the contraction of the covariant
acceleration with the dual Weyl tensor vanishes. This possibility
yields the following set of PDEs
\be
\theta_{,t\Theta} = \theta_{,t\phi} = 
\frac{\partial}{\partial r} \left(\frac{\theta_{,\Theta}}{r}\right)=
\frac{\partial}{\partial r} \left(\frac{\theta_{,\phi}}{r}\right)=0\,,
\ee
the solution of which is given by
\be
\label{possibletheta}
\theta = F(t,r) + r G(\Theta,\phi)\,.
\ee
Note that this scalar field possesses a non-vanishing covariant
acceleration, namely $v_{tt}$, $v_{tr}$, $v_{rr}$, $v_{\Theta\Theta}$,
$v_{\Theta \phi}$ and $v_{\phi \phi}$ are non-vanishing, e.g.~
\ba
v_{tt} &=& \partial_{rr} F -  \frac{M}{r^2} \left(1-\frac{2M}{r}\right) \left(\partial_r F  + G\right)\,.
\ea
For the choice of $\theta$ given in Eq.~\eqref{possibletheta} the
Schwarzschild solution is always a solution of the modified theory.
Note that Eq.~\eqref{possibletheta} reduces to the canonical choice
for $G = 0$ and $F = t/\mu$, for which the only non-vanishing
component of the covariant acceleration is $v_{tr} = - M/(r^2 f \mu)$.

This simple calculation of the most general form of the scalar field
that respects the Schwarzschild solution leads to two important
consequences:
\begin{itemize}
\item The existence of specific solutions depends sensitively on the
  choice of the scalar field.
\item The satisfaction of the Pontryagin constraint is a necessary but
  not a sufficient condition for the C-tensor to vanish.
\end{itemize}
In order to illustrate the second point, let us consider the scalar
field $\theta = m_{CS} \sin{\Theta}$, with $m_{CS}$ a constant. Then
the Pontryagin constraint is still satisfied, but the C-tensor has one
non-vanishing component,
\be
C_{t \phi} = \frac{3 M m_{CS}}{r^4} \sin^2{\Theta} \left(1 - \frac{2
    M}{r} \right)\,,
\ee
and the Schwarzschild line element [Eq.~\eqref{Schw}] is no longer a
solution to the modified field equations [Eq.~\eqref{eom}].

%------------------------------------------------------
\subsection{Spherically symmetric metrics}

Let us now pose the question whether there can be non-GR solutions in CS
modified gravity that preserve spherical symmetry. Any line element respecting this symmetry must be diffeomorphic to (cf.~e.g.~\cite{Balasin:2004gf})
\be
\label{spherical}
ds^2=g_{\alpha\beta}(x^\gamma)\,dx^\alpha dx^\beta + \Phi^2(x^\gamma)\, d\Omega^2_{{\cal S}^2}\,,
\ee
where $g_{\alpha\beta}(x^\gamma)$ is a Lorentzian 2-dimensional metric with some coordinates $x^\gamma$, $\Phi(x^\gamma)$ is a scalar field (often called ``dilaton'' or ``surface radius'') and $d\Omega^2_{{\cal S}^2}$ is a line element of the round 2-sphere, with some coordinates $x^i$.
For such a line element, one can show straightforwardly that
the Pontryagin constraint is always satisfied (cf.~e.g.~appendix A of
\cite{Grumiller:2002nm}), and that the only non-vanishing components
of the Ricci tensor are $R_{\alpha\beta}$ and $R_{ij}$. On the other hand, for the most
general scalar field $\theta$, the only non-vanishing components of
the C-tensor are of the form $C_{\alpha i}$. Remarkably, the C-tensor and the Ricci tensor decouple and
both have to vanish independently as a consequence of the modified
field equations. In other words, for spherically symmetric line
elements there cannot be solutions that live in ${\cal{CS}} \;
\backslash \; {\cal{P}}$. Instead all solutions are pushed to
${\cal{P}}$, which then uniquely leads to the Schwarzschild solution
by virtue of the Birkhoff theorem~\footnote{The persistence of Birkhoff's theorem in CS modified gravity was first proved by one of us (NY) in collaboration with C.~Sopuerta~\cite{nico:carlos}.}.

We have just shown that for all spherically symmetric situations the
vacuum solutions to the CS modified field equations live in
${\cal{P}}$, and therefore are given uniquely by the Schwarzschild
solution. For non-vacuum solutions with the same symmetries, similar
conclusions hold, since the field equations still decouple into
non-vacuum Einstein equations and the vanishing of the C-tensor.
Therefore, all solutions are again pushed to ${\cal{P}}$ and
spherically symmetric solutions of GR (such as the Reissner-Nordstr\"om BH or Friedmann-Robertson-Walker
spacetimes) persist in CS modified gravity, provided $\theta$ is of
the form
\be
\label{possibletheta2}
\theta = F(x^\gamma) + \Phi(x^\gamma)\, G(x^i)\,.
\ee
This result is completely analog to Eq.~\eqref{possibletheta}.
In all spherically symmetric scenarios, the solutions to the CS
modified field equations live in ${\cal{P}}$ and the expectations of
Sec.~\ref{se:sol} hold.

%------------------------------------------------------------------
\subsection{Losing the Kerr solution}

As an example of a relevant GR solution that does not persist in the
modified theory we consider the Kerr solution. The Kerr metric yields
a non-vanishing Pontryagin term~\footnote{Similar conclusions hold for
  the Kerr-Newman and Kerr-NUT spacetimes.}, which in Boyer-Lindquist
coordinates
\begin{multline}
\label{kerr}
ds^2=-\frac{\Delta-a^2\sin^2\Theta}{\Sigma}dt^2-\frac{4aMr\sin^2\Theta}{\Sigma}dtd\phi\\
+\frac{(r^2+a^2)^2-a^2\Delta\sin^2\Theta}{\Sigma}\sin^2\Theta d\phi^2+\frac{\Sigma}{\Delta}dr^2+\Sigma d\Theta^2
\end{multline}
can be written as
\ba
\label{Pkerr}
\pont &=& 96 \frac{a M^2r}{\Sigma^6}\cos{\Theta} \left(r^2 - 3 a^2
  \cos^2{\Theta} \right)
\nonumber  \\
&& \left(3 r^2 - a^2 \cos^2{\Theta} \right)\,,
\ea
with $\Sigma=r^2+a^2\cos^2\Theta$ and $\Delta=r^2+a^2-2Mr$. In light
of the physical interpretations of Sec.~\ref{se:pont}, one would
expect this result since the Kerr spacetime possesses a complex
Newman-Penrose scalar $\Psi_2$.

The Pontryagin constraint is satisfied in certain limits. For example,
as the Kerr parameter goes to zero, $a \to 0$, the Schwarzschild
solution is recovered and the right-hand side of Eq.~\eqref{Pkerr}
vanishes. Similarly, in the limit as the mass goes to zero, $M \to 0$,
the right-hand side of Eq.~\eqref{Pkerr} also vanishes.  However, for
any finite $a$ and $M$ the Pontryagin term is non-vanishing and, thus,
the Kerr spacetime cannot be a solution to the CS modified field
equations~\cite{Konno:2007ze}.

What line element then replaces the Kerr solution in the modified
theory? A reasonable attempt to construct a spinning BH in CS gravity
is to consider generic axisymmetric and either static or stationary
line elements, which we shall investigate in the next sections.

%%%%%%%%%%%%%%%%%%%%%%%%%%%%%%%%%%%%%%%%%%%%%%%%%%%%%%%%%%%%%%%%%%%%%%%%%%
\section{Static, axisymmetric solutions}
\label{static}

Before embarking on a {\emph{tour de force}} through generic
stationary and axisymmetric solutions, we shall first consider the
simpler case of static and axisymmetric solutions. Following
\cite{waldgeneral}, the most general static and axisymmetric line
element is diffeomorphic to
\be
\label{ds2}
ds^2 = -V dt^2 + V^{-1} \rho^2 d\phi^2 +
\Omega^2 \left(d\rho^2 + \Lambda dz^2\right),
\ee
where we have three undetermined functions of two coordinates:
$V(\rho,z)$, $\Omega(\rho,z)$ and $\Lambda(\rho,z)$. The two commuting
Killing fields, $\xi^a=\left(\partial_t\right)^a$ and
$\psi^a=\left(\partial_{\phi}\right)^a$, are associated with
stationarity and axisymmetry respectively. However, since there is no
cross-term $dtd\phi$, the line element of Eq.~\eqref{ds2} is not just
stationary but also static. The components of its Ricci tensor are
given by
\ba 
\label{Ricci-static}
R_{t\phi} &=&  R_{t\rho} = R_{tz} = R_{\phi\rho} = R_{\phi z} = 0\,,  \\
R_{tt}  &=& \frac{1}{2\Omega^2}\, \left[ 
\, V_{,\rho\rho}  
+\,\frac{V_{,zz}}{\Lambda} 
+\,\frac{V_{,\rho}}{\rho}
-\, \frac{V_{,\rho}^2}{V}
-\,\frac{V_{,z}^2}{V\Lambda}\right.\nonumber \\
&&\qquad \left.+ \frac{V_{,\rho}\Lambda_{,\rho}}{2\Lambda}  
-\frac{V_{,z}\Lambda_{,z}}{2\Lambda^2} \right]\,,
\label{Ricci-staticm1} \\
R_{\phi\phi} &=& \frac{1}{2 \Omega^2}\left[
\frac{\rho^2 V_{,\rho\rho}}{V^2} 
+\frac{\rho^2 V_{,zz}}{V^2 \Lambda}
+\frac{\rho V_{,\rho}}{V^2} 
-\frac{\rho^2 V_{,\rho}^2}{V^3}
-\frac{\rho^2 V_{,z}^{2}}{V^3\Lambda}
\right. \nonumber \\
&&\qquad\left. 
-\frac{\rho\Lambda_{,\rho}}{V \Lambda}  
+\frac{\rho^2 \Lambda_{,\rho} V_{,\rho}}{2V^2 \Lambda}
-\frac{\rho^2\Lambda_{,z} V_{,z}}{2 V^2} \right]\,, 
\\
R_{\rho\rho} &=&
\frac{V_{,\rho}}{\rho V}  
-\frac{V_{,\rho}^2}{2V^2}
-\frac{\Omega_{,\rho\rho}}{\Omega}
-\frac{\Omega_{,zz}}{\Lambda\Omega}
+\frac{\Omega_{,\rho}}{\rho\Omega} 
+\frac{\Omega_{,\rho}^2}{\Omega^2}
+\frac{\Omega_{,z}^2}{\Lambda\Omega^2}  
\nonumber \\
&&
-\frac{\Lambda_{,\rho\rho}}{2\Lambda}
+\frac{\Lambda_{,\rho}^2}{4\Lambda^2}
-\frac{\Omega_{,\rho}\Lambda_{,\rho}}{2\Lambda\Omega}
+\frac{\Omega_{,z} \Lambda_{,z}}{2\Lambda^2\Omega} 
\,,  
\\
R_{zz} &=&
-\frac{V_{,z}^2}{2V^2}
-\frac{\Lambda \Omega_{,\rho\rho}}{\Omega} 
-\frac{\Omega_{,zz}}{\Omega} 
-\frac{\Omega_{,\rho}\Lambda}{\rho\Omega}
+\frac{\Omega_{,\rho}^2\Lambda}{\Omega^2} 
+\frac{\Omega_{,z}^2}{\Omega^2} 
\nonumber \\
&&
-\frac{\Lambda_{,\rho\rho}}{2} 
-\frac{\Lambda_{,\rho}}{2\rho} 
+\frac{\Lambda_{,\rho}^2}{4\Lambda}
-\frac{\Omega_{,\rho} \Lambda_{,\rho}}{2\Omega}
+\frac{\Omega_{,z} \Lambda_{,z}}{2\Omega\Lambda}\,,
\label{eq:Ricci-staticm2} \\
R_{\rho z} &=& 
-\frac{V_{,z} V_{,\rho}}{2V^2} 
+\frac{V_{,z}}{2\rho V}
+\frac{\Omega_{,z}}{\rho\Omega}
\,, 
\label{eq:Ricci-static}
\ea 
and exhibit only five non-vanishing components.  With the canonical
choice of the CS scalar field [Eq.~\eqref{canon}] it is now
straightforward to check that the five corresponding components of the
C-tensor vanish,
\begin{equation}
C_{tt}=C_{\phi\phi}=C_{\rho\rho}=C_{zz}=C_{\rho z}=0 \,.
\end{equation}
As in the spherically symmetric case, we are faced with the remarkable
consequence that the field equations [Eq.~\eqref{eom}] decouple into
the vacuum Einstein equations plus the vanishing of the C-tensor, viz.
\be
\label{decoupled}
R_{ab}=0\,,\qquad C_{ab}=0\,.
\ee
In other words, using the classification of Sec.~\ref{se:sol}, all
static and axisymmetric solutions live in ${\cal{P}}$, which again
confirms previous expectations.

With these considerations in mind, we can now simplify the line
element of Eq.~\eqref{ds2}. From \cite{waldgeneral}, the function
$\Lambda$ can be chosen to be constant, e.g.~$\Lambda=1$, and
therefore the line-element reduces to the Weyl class,
\be
\label{dsWeyl}
ds^2=-e^{2U}dt^2+e^{-2U}\left[e^{2k}(d\rho^2+dz^2)+\rho^2d\phi^2\right]\,.
\ee
The vacuum Einstein equations then simplify to
\be
\label{EE-cyl}
\Delta U=0\,,\quad k_{,\rho} = \rho (U_{,\rho}^2-U_{,z}^2)\,,\,\,
k_{,z}=2\rho U_{,\rho} U_{,z}\,, 
\ee
where $\Delta = \partial^2/\partial\rho^2+1/\rho
\partial/\partial_\rho+\partial^2/\partial z^2$ is the flat space
Laplacian in cylindrical coordinates. The function $U$ thus solves a
Laplace equation, and for any such solution the function $k$ can be
determined by a line integral~\cite{Stephani:2003tm}.

The Pontryagin constraint is fulfilled automatically for all line
elements diffeomorphic to Eq.~\eqref{dsWeyl}, but as we have seen in
the previous sections, this is not sufficient to achieve $C_{ab}=0$.
For example, with the choices \cite{Chazy:1924,Curzon:1924} ($m$ is constant)
\be
\label{chazy}
U=-\frac {m}{\sqrt{\rho^2+z^2}}\,,\qquad k = -\frac{m^2\rho^2}{2(\rho^2+z^2)^2}\,,
\ee
the vacuum Einstein equations hold and the Pontryagin constraint is
fulfilled, but the C-tensor has the non-vanishing components
\be
C_{\rho \phi} = \frac{2 m^4 \rho^3 z}{\mu\left(\rho^2 +
    z^2\right)^5} \exp\left[\frac{m^2 \rho^2}{\left(\rho^2 +
      z^2\right)^2} \right], 
\quad
C_{z \phi} = \frac{1}{2 \rho z} C_{\rho \phi}\,. 
\ee

Since the C-tensor must vanish independently from the Einstein
equations, once more we are faced with three distinct possibilities,
identical to those described in Sec.~\ref{se:sol}. The first
possibility (option $1$ in Sec.~\ref{se:sol}) is to demand that the
Weyl tensor vanishes, but since also the Ricci tensor vanishes, the
spacetime would have to be flat. The second possibility (option $2$ in
Sec.~\ref{se:sol}) is to demand that the covariant acceleration of
$\theta$ vanishes, {\emph{i.e.}}, $\nabla_a\theta$ is a covariantly
constant vector. However, as we have mentioned already, a vacuum
solution with a covariantly constant vector field must be either flat,
or the vector must be a null-vector.  The first alternative is
trivial, while the second one is not particularly interesting in the
context of static axisymmetric spacetimes. We shall discuss the latter
possibility further in Sec.~\ref{se:KCS}.

The only remaining possibility (option $3$ in Sec.~\ref{se:sol}) is
for the contraction of the covariant acceleration and the dual Weyl
tensor to vanish. The C-tensor can then be simplified to
\be
C^{ab} \propto \Gamma^t{}_{\rho t}{\,^\ast\!}C^{t(ab)\rho} +
\Gamma^t{}_{zt}{\,^\ast\!}C^{t(ab)z} = 0\,,
\ee 
which has only two non-vanishing components. Using the Einstein
equations to simplify these expressions we obtain a set of nonlinear
PDEs,
\ba
U_{,\rho z} U_{,\rho} + U_{,zz} U_{,z} &=& 2 \rho \left( U_{,z}^3
  U_{,\rho} + U_{,\rho}^3 U_{,z} \right) - \frac{2}{\rho} U_{,\rho}
U_{,z}\,,
\nonumber \\
U_{,\rho z} U_{,z} - U_{,zz} U_{,\rho} &=& \rho \left( U_{,\rho}^4 -
  U_{,z}^4 \right) - \frac{1}{\rho} U_{,\rho}^2\,.
\ea
We used Maple to obtain some solutions to these PDEs. The Schwarzschild
solution
\be U=\frac12
\ln{\frac{\sqrt{\rho^2+(z+M)^2}+\sqrt{\rho^2+(z-M)^2}-2M}{\sqrt{\rho^2+(z+M)^2}+\sqrt{\rho^2+(z-M)^2}+2M}}
\label{eq:SWeyl}
\ee
of course solves these PDEs. Some other simple solutions are $U=U_0$,
$U = U_0 \pm \ln \rho$ and $U=U_0+\ln{(\sqrt{\rho^2+z^2}+z)}/2$, where
$U_0$ is a constant. Not only do these solution yield a vanishing
Ricci tensor, but they also yield a vanishing Riemann tensor, which
shows they are Minkowski spacetime in disguise. In addition to these
trivial solutions, there exist exactly two more:
\be
\label{eq:Per1}
ds^2 = -\frac 1z dt^2 + z dz^2 + z^2(d\rho^2+\rho^2d\phi^2)
\ee
and
\ba
ds^2 &=& -\left(\frac{2m}{z}-1\right)dt^2 +
\left(\frac{2m}{z}-1\right)^{-1} dz^2 
\nonumber \\ 
&+& z^2(d\rho^2+\sinh^2{\!\rho}\, d\phi^2)
\label{eq:Per2}
\ea
While these solutions certainly are non-trivial, neither the first
[Eq.~\eqref{eq:Per1}] nor the second [Eq.~\eqref{eq:Per2}] solution is
physically relevant. The former has a naked singularity at $z=0$,
while the latter, whose singularity at $z=0$ is screened by a Killing
horizon at $z=2m=\rm const.$, possesses a Killing vector $k^a=(\partial_t)^a$ 
that is spacelike in the ``outside'' region $z>2m$,
{\emph{i.e.}}~$g_{ab}k^a k^b=1-2m/z > 0$.

Let us now prove that these are the only solutions to the modified
field equations. The crucial observation is that the spatial part of
the C-tensor for static, axisymmetric spacetimes reduces to the
$3$-dimensional Cotton tensor~\cite{Jackiw:2003pm}. This tensor
vanishes if and only if the corresponding $3$-dimensional space
(spatial sector of the $4$-dimensional metric) is conformally flat,
{\emph{i.e.}}~$g_{i j} = \Phi \delta_{i j}$, where the conformal
factor $\Phi$ is a function of the coordinates and $\delta_{ij}$ is
the spatial part of the Minkowski metric. We may then exploit a result
by Luk{\'a}cs and Perj{\'e}s \cite{Lukacs:1982} that the line elements
of Eqs.~\eqref{eq:Per1},~\eqref{eq:Per2} and~\eqref{Schw} are the only
static and axisymmetric solutions that are spatially conformally flat.
Therefore, it follows that these equations are the only solutions to
the modified field equations.

The above considerations also apply to more general CS scalar fields. All
simplifications hinge on the decoupling of the modified field
equations, which occurs if and only if $\theta_{,t\phi}=0$. We can
solve this PDE to obtain
\be
\label{v-gen-static}
\theta = \theta_1(t,\rho,z) + \theta_2(\rho,z,\phi).
\ee
For all scalar fields of this form, the
modified field equations decouple and the C-tensor has five
non-vanishing components, which define a system of PDEs for one
of the two arbitrary functions $k$ or $U$. However, we do not expect
more solutions to arise in this way, since this case leads to the same
constraints as the canonical one, plus three extra PDEs, which
essentially compensate the freedom to tinker with the two arbitrary
functions in Eq.~\eqref{v-gen-static}.

The most general CS scalar field, however, does not allow for a
decoupling of the type described above. If the scalar field has
$\theta_{,t\phi} \neq 0$, then the $(\rho,\rho)$, $(\rho,z)$ and
$(z,z)$ components of the modified field equations do not decouple.
However, the $(t,t)$ and $(\phi,\phi)$ components still do decouple
because the corresponding C-tensor components vanish. The equation
\be
R^t{}_t + R^{\phi}{}_{\phi} = -\frac{1}{2} \frac{\Lambda_{,\rho}}{\Lambda
  \rho^2 \Omega^2} = 0,
\ee
forces $\Lambda$ to be a function of $z$ alone. Through a
diffeomorphism, this function can be set to unity, as argued in
\cite{waldgeneral}.

The modified field equations are too difficult to solve analytically
with Maple, so in order to study solutions that do not lead to a
decoupling of the modified field equations, we shall assume for
simplicity $\theta = \tilde{\theta}(t,\phi)$.  From the Ricci sector
of the field equations ($R_{tt} = 0 = R_{\phi \phi}$) we find that $U$
is again a solution of $\Delta U = 0$.  We can use this relation to
simplify the C-tensor, and the ensuing equations $C_{t \phi} =
C_{t\rho} = C_{t z} = C_{\phi\rho} = C_{\phi z} = 0$ lead to a system
of second order PDEs for $\theta$ and $k$. We investigated this system
with Maple and found that solutions exist if and only if $\theta$ is a
function of only one variable, {\emph{i.e.}}~$\theta = \theta(t)$ or
$\theta = \theta(\phi)$.  These results indicate that there are no
solutions of the modified field equations if $\theta$ is bivariate.

In summary, we have shown in this section that the field equations
decouple if the CS scalar field solves $\theta_{,t\phi} = 0$, and
their solution is the Schwarzschild BH and two additional (unphysical)
solutions [Eqs.~\eqref{eq:Per1} and~\eqref{eq:Per2}]. For CS fields
that satisfy $\theta_{,t\phi} \neq 0$, the modified field equations do
not seem to have a solution. Therefore, there are no static and
axisymmetric solutions in CS gravity, apart from the Schwarzschild BH
and some unphysical solutions, irrespective of the CS scalar field.

%%%%%%%%%%%%%%%%%%%%%%%%%%%%%%%%%%%%%%%%%%%%%%%%%%%%%%%%%%%%%%%%%%%%%%%%%%%%%%
\section{Stationary, axisymmetric solutions}
\label{stationary}

%-----------------------------------------------------------------------------
\subsection{General line elements}
\label{gen-metric-st-axi}

Equipped with the tools from the previous section, we drop
the requirement of staticity and replace it by the weaker one of
stationarity. In essence, this means that we shall allow the
gravitomagnetic sector of the metric to be different from zero.  The
most general, stationary and axisymmetric line-element is
diffeomorphic to \cite{Stephani:2003tm}
\be
ds^2 = -V \left(dt - w d\phi\right)^2 + V^{-1} \rho^2 d\phi^2 +
\Omega^2 \left(d\rho^2 + \Lambda dz^2 \right)\,,
\ee
where the functions $V$, $w$, $\Omega$ and $\Lambda$ depend on
$\rho$ and $z$, only. This line element is identical to Eq.~\eqref{ds2} as
$w \to 0$. In GR, the function $w$ can be identified with the angular
velocity.
The Ricci tensor for this line element is similar to $R_{ab}^{\rm static}$
[Eqs.~\eqref{Ricci-static}-\eqref{eq:Ricci-static}] and its components
are
\begin{align}
R_{t\rho} &= R_{tz} = R_{\phi \rho} =  R_{\phi z} = 0\,,\label{eq:Ricci-stationary}\\
R_{tt} &= R_{tt}^{\rm static} + \Big( w_{,\rho}^2 + \frac{w_{,z}^2}{\Lambda} \Big) \frac{V^3}{2 \rho^2 \Omega^2} \,,\\
\label{eq:Ricci-stationarym}
R_{t \phi} &= \frac{w}{2 \Omega^2} \Big[-V_{,\rho\rho}-\frac{V_{,zz}}{\Lambda}-\frac{V_{,\rho}}{\rho}+\frac{V_{,\rho}^2}{V}+\frac{V_{,z}^2}{V \Lambda}  
\nonumber \\
&- 2 \frac{V_{,\rho} w_{,\rho}}{w}  - \left.  2 \frac{V_{,z}w_{,z}}{w \Lambda}-\frac{w_{,\rho \rho}V}{w}  - \frac{w_{,zz} V}{w \Lambda} 
\right. 
\nonumber \\
&+ \left. \frac{w_{,\rho}V}{\rho w}-\frac{w_{,\rho}^2V^3}{\rho^2}-\frac{w_{,z}^2V^3}{\rho^2 \Lambda}  
\right. 
\nonumber \\
&- \frac{\Lambda_{,\rho} V_{,\rho}}{2 \Lambda}+\frac{\Lambda_{,z} V_{,z}}{2\Lambda^2}-\frac{\Lambda_{,\rho}w_{,\rho}V}{2 w\Lambda}+ \frac{\Lambda_{,z}w_{,z}V}{2 w \Lambda^2}\Big]\,,\\
R_{\rho \rho} &= R_{\rho \rho}^{\rm static} + w_{,\rho}^2 \frac{V^2}{2 \rho^2}\,,\\
R_{zz} &= R_{zz}^{\rm static} + w_{,z}^2 \frac{V^2}{2 \rho^2}\,,\\
R_{\rho z} &= R_{\rho z}^{\rm static} + w_{,\rho}w_{,z}\frac{V^2}{2 \rho^2} \,,\\
R &= R^{\rm static} +  \Big(w_{,\rho}^2 + \frac{1}{\Lambda} w_{,z}^2  \Big)\frac{V^2}{2 \rho^2 \Omega^2}\,.
\end{align}
The somewhat lengthy component $R_{\phi\phi}$ can be deduced from $R$ and the other components. The quantity $R^{\rm static}$, 
\begin{multline}
R^{\rm static} = 
\frac{1}{2\Omega^2} \Big[ 
2\frac{V_\rho}{V\Lambda}
-\frac{V_\rho^2}{V^2} 
-\frac{V_z^2}{V^2\Lambda} \\
-4\frac{\Omega_{\rho\rho}}{\Omega}
-4\frac{\Omega_{zz}}{\Lambda\Omega} 
+4\frac{\Omega_\rho^2}{\Omega^2} 
+4\frac{\Omega_z^2}{\Lambda\Omega^2} \\
-2\frac{\Lambda_{\rho\rho}}{\Lambda}
-2\frac{\Lambda_\rho}{\rho\Lambda} 
+\frac{\Lambda_\rho^2}{\Lambda^2}
-2\frac{\Lambda_\rho\Omega_\rho}{\Lambda\Omega}     
+2\frac{\Lambda_z\Omega_z}{\Lambda^2\Omega}   \Big]\,,
\label{eq:Ricciscalar-stationary}
\end{multline}
is the trace of Eqs.~\eqref{Ricci-static}-\eqref{eq:Ricci-static}.

As before, let us begin with the canonical choice for the CS scalar
field, namely Eq.~\eqref{canon}. Then
the only non-zero components of the C-tensor are $C_{\rho t}$, $C_{z
  t}$, $C_{\rho\phi}$ and $C_{z\phi}$. As in the previous cases, there is a
decoupling of the field equations that allows us to set $\Lambda = 1$
and to consider the slightly simpler line element
(Lewis-Papapetrou-Weyl metric)
\be
\label{LPW}
ds^2=-e^{2U}(dt-wd\phi)^2+e^{-2U}\left[e^{2k}(d\rho^2+dz^2)+\rho^2d\phi^2\right] \,,
\ee 
where again the functions $U$, $k$ and $w$ depend on the coordinates
$\rho$ and $z$ only. With this line element, the last lines vanish in
the multi-line expressions for the Ricci tensor,
Eqs.~\eqref{Ricci-staticm1}-\eqref{eq:Ricci-staticm2},
\eqref{eq:Ricci-stationarym} and \eqref{eq:Ricciscalar-stationary},
because $\Lambda=1$. The vacuum Einstein
equations simplify considerably with $\Lambda=1$. Essentially, they are similar to
Eq.~(\ref{EE-cyl}) but with a complicated source and an additional
equation for $w$. Even within GR, explicit solution to this set of
PDEs can only be found in certain special
cases~\cite{Stephani:2003tm}.

The Pontryagin constraint for the line element of Eq.~(\ref{LPW}) is
not satisfied in general. This constraint yields a complicated second
order PDE for $w$, $U$ and $k$, presented in appendix \ref{app:A},
which of course is trivially satisfied as $w \to 0$. Certain solutions
to the PDE in appendix \ref{app:A} can be obtained, e.g. ($\bar{w} :=
e^{2U} w$)
\begin{subequations}
\label{pont}
\begin{align}
k &= k(\rho,z)\, & \bar{w} &= c_1 e^{2U}\,, &  U &= U(\rho,z)\,,
\\
k &= k(\rho,z)\, & \bar{w} &= \pm\rho \,, &  U &= U(\rho,z)\,,
\\
k &= \ln(\rho) + \tilde{k}(z)\,, & \bar{w} &= \tilde{w}(z) \rho\,, & U &=
\frac{1}{2} \ln(\rho) + c_1\,,
\\
k &= k(\rho)\,, & \bar{w} &= \bar{w}(\rho)\,, & U &= U(\rho)\,,
\end{align}
\end{subequations}
where $c_1$ is a constant. The first line reduces to static solutions
upon redefining $t'=t-c_1\phi$. The second line leads to metrics of
Petrov type $II$, the so-called van Stockum class, which we shall
discuss in Sec.~\ref{se:vS}. The third line of Eq.~(\ref{pont}) cannot
be made to solve the modified field equations.  The last line implies
cylindrical symmetry, which again via the field equations leads to
flat spacetime. We have thus been unable to find non-trivial solutions
either by hand or using symbolic manipulation software~\footnote{We
  were able to find additional solutions with Maple, but upon imposing
  the field equations they reduced to previously studied or trivial
  spacetimes.}.

Unlike the previous section, we cannot provide here a truly exhaustive
discussion of all solutions of the decoupled field equations. This is
because $C_{ab}=0$ does not necessarily imply spatial conformal
flatness for the stationary case. Based on the evidence found so far,
it seems unlikely that there are other non-trivial and physically
interesting solutions besides the static ones. This is because the
vacuum Einstein equations [$R_{ab}=0$] already determine the function
$k$ uniquely up to an integration constant, and also impose strong
restrictions on the functions $U$ and $w$~\cite{Stephani:2003tm}. The
constraints $C_{ab}=0$ impose four additional conditions on these
functions that can be found in~\cite{nicoswebsite}. Since the system
of partial differential equations is over-constrained, it is unlikely
that additional solutions exist. Therefore, whenever the field
equations decouple into $R_{ab}=0=C_{ab}$ we do not expect physically
relevant solutions besides the Schwarzschild one and its flat space
limit.

The decoupling exhibited by the modified field equations does not
occur only for the canonical choice of the CS scalar field. In order for
such a decoupling to occur, the following system of PDEs must be
satisfied:
\be
\theta_{,tt} = 
\theta_{,\phi\phi} = 
\theta_{,t\phi} =
\theta_{,\rho} =
\theta_{,z} = 0\,,
\ee
which yields the solution
\be
\label{stationary-axi-theta}
\theta = \frac{t}{\mu} + \frac{\phi}{\nu}\,,
\ee
with constant $\mu$, $\nu$. The canonical choice is
recovered as $\nu \to \infty$. 

But what if the scalar field is not of the form of
Eq.~(\ref{stationary-axi-theta})? In this case, the field equations do
not decouple and solving the entire system is much more complicated.
However, we can deduce from Eq.~\eqref{eq:Ricci-stationary} that still
the four C-tensor components $C_{\rho t}$, $C_{z t}$, $C_{\rho\phi}$
and $C_{z\phi}$ have to vanish. Therefore, even though no decoupling
occurs, the same issue of an over-constrained system of equations does
arise, analogous to the one in Sec.~\ref{gen-metric-st-axi}. Even with
this generalization, it is still quite difficult to find solutions to
the coupled system. In general, one might be able to find solutions
both of class ${\cal{P}}$ and class ${\cal{CS}} \; \backslash \;
{\cal{P}}$ because non-canonical CS fields allow for general $\theta$,
which entails a new degree of freedom. We shall see in
Sec.~\ref{se:vS} that for a simplified subclass of stationary and
axisymmetric line elements, which automatically satisfy the Pontryagin
constraint, solutions can indeed be found, including mathematical BHs.

%-----------------------------------------------------------------------------
\subsection{Van Stockum line element}\label{se:vS}

We study now a slightly less general line element that still is
stationary and axisymmetric, namely the van Stockum line
element~\cite{Stephani:2003tm}
\be
ds^2 = \rho \Omega dt^2 - 2 \rho dt d\phi + \frac{1}{\sqrt{\rho}}
\left(d\rho^2 + dz^2\right)\,,
\label{eq:vSle}
\ee
where the only arbitrary function is $\Omega = \Omega(\rho,z)$. The
metric is different from that considered in Eq.~\eqref{LPW} since it
does not possess a $d\phi^2$ component. The only non-vanishing
component of the Ricci tensor for such a spacetime is
\be
R_{tt} = -\frac{\rho^{3/2}}{2} \Delta \Omega\,, \label{eq:RttvS}
\ee
where again $\Delta$ is the flat space Laplacian in cylindrical
coordinates. 

The Pontryagin constraint is automatically satisfied for the van
Stockum line element even though it is of Petrov type $II$, precisely
because of the vanishing $d\phi^2$ term. The $tt$ component of the
modified equations then determines $\Omega$, and this forces all other
components of the C-tensor to vanish, except for $C_{t\phi}$ and
$C_{\phi \phi}$ that are automatically zero. These constraints act as
a system of PDEs for the scalar field, whose unique solution is
$\theta = \theta(\rho,z)$. Note that the canonical choice for $\theta$
is not compatible with the van Stockum line element. The remaining PDE
$R_{tt} + C_{tt} = 0$ can be solved for $\theta$ and $\Omega$, where
$C_{tt}$ now simplifies to
\begin{align}
C_{tt} &= \frac{\rho^2}{2}\Big[(\theta_{,\rho\rho}-\theta_{,zz})\big(\Omega_{,\rho z}+\frac{3}{4\rho}\Omega_{,z}\big) \nonumber \\
& + \theta_{,\rho} \big(\Omega_{,zzz} + \Omega_{,\rho\rho z} + \frac{3}{2\rho} \Omega_{,\rho z} + \frac{3}{8\rho^2}\Omega_{,z} \big) \nonumber \\
& - \theta_{,z} \big(\Omega_{,\rho\rho\rho} + \Omega_{,\rho zz} + \frac{9}{4\rho} \Omega_{,\rho \rho} + \frac{3}{4\rho} \Omega_{,zz} + \frac{3}{8\rho^2} \Omega_{,\rho} \big) \nonumber \\
& - \theta_{,\rho z} \big(\Omega_{,\rho\rho} - \Omega_{,zz} + \frac{3}{2\rho} \Omega_{,\rho} \big) \Big]\,.
\end{align}
Combining this with $R_{tt}$ from Eq.~\eqref{eq:RttvS} we find two
simple solutions of Eq.~\eqref{eom}:
\be
\label{not-flat}
\Omega = c, \qquad \theta = \theta(\rho,z),
\ee
where $c$ is a constant and
\be
\label{not-GR}
\Omega = c + \frac{d}{\sqrt{\rho}}, \qquad \theta = \frac{2}{3} \sqrt{\rho}\, z + \tilde{\theta}(\rho),
\ee
where $c$ and $d$ are constants \footnote{There is a straightforward
  generalization of Eq.~\eqref{not-GR}, namely
  $\Omega=c+d/\rho^\alpha$, $\theta = \beta \sqrt{\rho}\, z$, where
  $\alpha$ is an arbitrary constant and $\beta = 8\alpha/(8\alpha^2 +2\alpha+3)$.}.  Equation~\eqref{not-flat} leads to zero Ricci and
C-tensor separately and it is thus a GR solution that belongs to the
subspace ${\cal{P}}$.  The ensuing metric is exceptional in that it
has a third Killing vector,
$t\partial_t-\phi\partial_\phi+ct\partial_\phi$.  Some of the
non-vanishing Riemann tensor components for this geometry are
\be
R_{t\rho t\rho} = \frac{c}{8 \rho}\,,\quad
R_{t\rho \rho \phi} = \frac{1}{8 \rho}\,,\quad
R_{t\phi t\phi} = \frac{1}{4} \sqrt{\rho}\,.
\ee
On the other hand, Eq.~\eqref{not-GR} is perhaps even more interesting
since it is not Ricci-flat, but has one non-vanishing component of the Ricci tensor, 
\be
R_{tt} = -\frac{d}{8 \rho} = - C_{tt}\,.
\ee
This solution is thus a non-GR solution and it
belongs to the subspace ${\cal{CS}} \; \backslash \; {\cal{P}}$. Some of the
non-vanishing components of the Riemann tensor for this solution are 
\be
R_{t \rho t \rho} = \frac{d+2c\sqrt{\rho}}{16 \rho^{3/2}}\,,\quad
R_{t \rho \rho \phi} = \frac{1}{8 \rho}\,,\quad
R_{t \phi t \phi} = \frac{1}{4} \sqrt{\rho}\,.
\ee
Notice that such a solution can represent a mathematical BH, provided
$\Omega$ vanishes for some $\rho$, {\emph{i.e.}}~a Killing horizon
emerges. We call these configurations ``mathematical BHs'' because
they are physically not very relevant: the Killing vector generating
axial symmetry is light-like, as evident from \eqref{eq:vSle}, and the
spacetime admits closed timelike curves which are not screened by a horizon \cite{Stephani:2003tm}. For $c=1$ and $d=-2m$ we recover \eqref{eq:conclusion1}.

Let us then summarize the most important conclusions of this section.
We have investigated stationary and axisymmetric solutions to the
modified field equations. We found that, for the canonical choice of
$\theta$, it is unlikely that solutions can be found that differ from
Minkowski and Schwarzschild. Nonetheless, for non-canonical choices of
this scalar, solutions must exist. This conclusion derives from the
investigation of a slightly less general stationary and axisymmetric
metric, namely that of van Stockum. For this line element we found a
solution to the modified field equations that lives in ${\cal{P}}$ and
a family of solutions that live in ${\cal{CS}} \; \backslash \;
{\cal{P}}$, both with non-canonical CS scalar fields. To our
knowledge, this is the first time an exact non-GR solution is found
for CS modified gravity, which in particular can represent
mathematical BH configurations.

%%%%%%%%%%%%%%%%%%%%%%%%%%%%%%%%%%%%%%%%%%%%%%%%%%%%%%%%%%%%%%%%%%%%%%%%%%%%%%
\section{Beyond the canon}
\label{beyond}

We have failed in finding an exact, stationary and axisymmetric
solution to the CS modified field equations representing a physical
spinning BH. A solution, however, already exists for a similar line
element, albeit in a perturbative sense. In~\cite{Alexander:2007vt}
and later in~\cite{Smith:2007jm}, a far-field solution to the CS
modified field equations with a canonical CS scalar field was found in
the weak-field/slow-motion approximation. This solution is identical
to the far-field expansion of the Kerr solution, except for the
addition of two new components in the gravitomagnetic sector of the
metric $g_{0i}$. These components vanish in GR, since only one
component is required and it is aligned with the angular momentum of
the spinning source. In CS gravity, the remaining components of
$g_{0i}$ are proportional to the curl of the spin angular momentum,
thus breaking axisymmetry but preserving stationarity. Such a
stationary, but non-axisymmetric BH will not emit gravitational waves,
but it will possess a non-trivial multipolar structure, with probably
more than just two non-vanishing multipoles. Such a far-field
structure suggests that perhaps the only way to obtain an analog to
the Kerr solution in CS gravity is to relax either the assumption of
axisymmetry or stationarity.  Alternatively, the van Stockum example
suggests that another possibility is to allow for a general CS scalar
field. In this case, however, the line element must significantly
differ from the Kerr metric such that it satisfies the Pontryagin
constraint~\footnote{We attempted to solve the modified field
  equations with a line element equal to Kerr plus a stationary and
  axisymmetric contribution and an arbitrary CS scalar field. No
  solution could be found exactly, while perturbatively such an Ansatz
  leads to non-asymptotically flat solutions~\cite{Konno:2007ze}.}.
We shall explore these possibilities in this section.

%------------------------------------------------------------------
\subsection{Killing embedding}\label{se:KCS}

We study now the possibility that the 'embedding coordinate', {\emph{i.e.}},
the velocity of the CS scalar field $\theta$, is a Killing vector.
Then, $v_a$ is covariantly conserved because of the Killing equation
($\nabla_{(a}v_{b)}=0$) and the fact that the connection is
torsion-free ($\nabla_{[a}v_{b]}=\nabla_{[a}\nabla_{b]}\theta=0$).
This puts a strong restriction on spacetime, which for a time-like
$v_a$ yields line elements that are diffeomorphic to
\be
ds^2=-dt^2+g_{ij}(x^k)dx^idx^j\,,
\label{eq:lala}
\ee
where $i$ and $j$ range over all coordinates except time.  Actually,
Eq.~\eqref{eq:lala} describes a special class of static spacetimes.
When studying static solutions to the CS modified field equations with
timelike $v_a$ in Sec.~\ref{static}, we found no physically relevant
solution besides Schwarzschild.  The same conclusions hold here,
except that we do not even recover Schwarzschild, so this route is not
a promising one. A similar discussion applies to spacelike Killing
vectors.

A more interesting situation arises if the vector field $v^a$ is a
null Killing vector, $v^a v_a = \nabla_{(a} v_{b)} = 0$.  In this
case, we get in an adapted coordinate system the line element
\be
\label{null-Killing}
ds^2 = -2 dv dx^1 + g_{ij}(x^k)dx^idx^j 
\ee
Once again, the Pontryagin constraint is immediately satisfied, the
Ricci tensor has non-vanishing $R_{ij}$ components, but no components
of the C-tensor vanish. Even when we pick a simple null Killing
embedding, e.g.~$v_a = \left(0,\chi,0,0\right)$ with $\chi =
{\rm{const.}}$, the C-tensor has complicated spatial non-vanishing
components and the modified field equations are too difficult to solve
in full generality. Therefore, we focus instead on an interesting
special case in the next subsection.

%-----------------------------------------------------------------
\subsection{pp-waves and boosted black holes}\label{se:pp}

As suggested at the end of Sec.~\ref{se:pont}, it might be possible to
find solutions to the modified field equations if one considers line
elements that represent exact gravitational wave solutions (pp-waves
\cite{Jordan:1960}). The line element for these waves is
\be
\label{pp}
ds^2=-2dvdu-H(u,x,y)du^2+dx^2+dy^2,
\ee
which is simply a special case of the line elements considered in the
previous subsection [Eq.~\eqref{null-Killing}]. Particular examples
of physical scenarios that are well-represented by Eq.~\eqref{pp} are
the Aichelburg-Sexl limits \cite{Aichelburg:1971dh} of various BHs. In
essence, this limit is an ultrarelativistic boost that keeps the
energy of the BH finite by taking a limit where its mass vanishes
while the boost velocity approaches the speed of light.  In
particular, Eq.~\eqref{pp} can be used to represent ultrarelativistic
boosts of the Kerr BH \cite{Lousto:1992th,Balasin:1995tb}.

Is it conceivable that a Kerr BH that moves ultrarelativistically
solves the modified field equations, even though the Kerr BH does not?
One of the main problems with the Kerr metric is that it does not
satisfy the Pontryagin constraint, cf.~Eq.~\eqref{Pkerr}, but that
constraint is trivially satisfied as $M\to 0$. Nonetheless, the
satisfaction of the Pontryagin constraint is only a necessary
condition, but not a sufficient one, to guarantee that the modified
field equations are also satisfied.   

In order to study this issue, let us find the appropriate expressions
for the Ricci and C-tensors. The only non-vanishing component of the
Ricci tensor for the line element of Eq.~\eqref{pp} is given by
\be
R_{uu}=\Delta H\,,\qquad \Delta:=\frac{\partial^2}{\partial
  x^2}+\frac{\partial^2}{\partial y^2}\,. 
\ee
In general, the components $C_{ux}$, $C_{uy}$, $C_{xx}$, $C_{yy}$,
$C_{xy}$ are all non-vanishing and form a system of PDEs for
$H$ and $\theta$. The $C_{xx}$, $C_{yy}$ and $C_{xy}$ components are
given by
\be
C_{yy} = -C_{xx} = \theta_{,vv} H_{,xy}\,,\quad
C_{xy} = \frac{1}{2} \theta_{,vv} \left(H_{,xx} - H_{,yy}\right)\,.
\ee

Let us first look for GR-solutions of class ${\cal{P}}$, such that
$R_{ab} = 0$ and $C_{ab} = 0$ independently. Since $C_{ab} = 0$, there
are two possibilities here: either $\theta_{,vv} = 0$ or $H_{,xy} = 0
= H_{,xx} - H_{,yy}$. In the latter case, $H$ is constrained to
\be
\label{sub_H}
H = \frac{1}{2} \left(x^2 + y^2\right) A(u) + x\,B(u) + y\,C(u) + D(u)\,,
\ee
which also forces $C_{ux}$ and $C_{uy}$ to vanish. The only component
of the field equations left is $(u,u)$, which upon simplification with
Eq.~\eqref{sub_H} yields $C_{uu} = 0$ and $R_{uu} = 2 A(u)$, so that
$A(u) = 0$. We have then found the solution
\be
H = x\, B(u) + y\, C(u)  + D(u) \,, \quad \theta = \theta(u,v,x,y)\,,
\ee
to the modified field equations. However, this solution is nothing
but flat space in disguise.

Another possibility to find GR-solutions is to pick $\theta$ such that
$C_{xx}$, $C_{yy}$ and $C_{xy}$ vanish, {\emph{i.e.}} $\theta_{,vv} =
0$. This condition leads to
\be
\label{red_theta}
\theta=\lambda(u,x,y) v + \tilde{\theta}(u,x,y)\,,
\ee
The remaining non-$(u,u)$ components of the C-tensor lead to 
\ba
C_{ux} &=& 0 \quad \rightarrow \quad  \lambda_{,x} H_{,xy} = \lambda_{,y} H_{,yy}
 \\
C_{uy} &=& 0 \quad \rightarrow \quad  \lambda_{,x} H_{,yy} = \lambda_{,y} H_{,xx}
\ea
where we have used $R_{uu} = 0$. The solution to this system of PDEs
leads either to flat spacetime or to
\be
\label{new-lambda}
\lambda(u,x,y) = \lambda(u).
\ee
Choosing Eq.~\eqref{new-lambda}, the remaining modified field equation
[the $(u,u)$ component] becomes
\ba
\label{lift}
\Delta H &=& 0\,, 
\\
\label{generation}
2 H_{,yy} \tilde\theta_{,xy} &=& 
H_{,xy} (\tilde\theta_{,yy}-\tilde\theta_{,xx})\,.
\ea
For some $H$ that solves the Einstein equations [{\emph{i.e.}}~the
Laplace equation in Eq.~\eqref{lift}], the C-tensor yields a PDE for
$\tilde{\theta}$ [Eq.~\eqref{generation}].  Thus we conclude that we
can lift any pp-wave solution of the vacuum Einstein equations to a
pp-wave solution of CS modified gravity (of class ${\cal{P}}$) by
choosing $\theta$ such that
Eqs.~\eqref{red_theta},~\eqref{new-lambda}-\eqref{generation} hold.

Let us give an example of this method to generate CS solutions by
studying ultrarelativistically boosted Kerr BHs, for which 
\be
\label{boosted-Kerr}
H = h_0 \delta(u) \ln\left(x^2 + y^2\right)\,
\ee
satisfies Eq.~(\ref{lift}). In Eq.~(\ref{boosted-Kerr}), $\delta(u)$
is the Dirac delta function and $h_0$ is a constant.  Inserting this
$H$ into Eq.~\eqref{generation} we find
\be
\label{tildetheta}
\tilde\theta = x \alpha\left(\frac{y}{x}\right) + \beta\left(x^2+y^2 \right)\,,
\ee
where $\alpha$ and $\beta$ are arbitrary functions of their respective
arguments $(y/x)$ and $(x^2+y^2)$.  Equation~\eqref{tildetheta},
together with Eq.~\eqref{red_theta} and~\eqref{new-lambda}, give the
full solution for the CS scalar field.  We have therefore lifted the
boosted Kerr BH to a solution of the modified field equations of class
${\cal{P}}$ by choosing the CS scalar field appropriately. For
$\theta=\lambda v$ we recover Eq.~\eqref{eq:intro17}.

Let us now search for non-GR solutions to the modified field
equations. Since all equations decouple except for the $(u,u)$
component, we must enforce that the non $(u,u)$-components of the
C-tensor vanish, {\emph{i.e.}~$\theta_{,v}=0$, which leads to
\be
\label{red_theta0}
\theta = \tilde{\theta}(u,x,y)\,.
\ee
%
%This is obviously more restrictive than Eq.~\eqref{red_theta} because
%we no longer can exploit Ricci-flatness. 
With Eq.~\eqref{red_theta0}, the only component of the modified field
equations left is again the $(u,u)$ one, which simplifies to a linear
third order PDE:
\begin{multline}
\label{ole}
(1+\tilde\theta_{,y}\partial_{,x}-\tilde\theta_{,x}\partial_{,y})\Delta  
H=\\=(\tilde\theta_{,xx}-\tilde\theta_{,yy})H_{,xy}-(H_{,xx}-H_{,yy})\tilde\theta_{,xy}\,.   
\end{multline}
For simplicity, we choose 
\be
\label{simplepptheta}
\tilde\theta=a(u)x+b(u)y+c(u)\,,
\ee
and Eq.~\eqref{ole} reduces to the Poisson equation
\be
\label{poisson}
\Delta H=f\,.
\ee
The source term $f$ solves a linear first order PDE
\be
bf_{,x}-af_{,y}-f=0\,,
\ee
whose general solution [assuming $b(u)\neq 0$]
\be
\label{f}
f(u,x,y)=e^{x/b(u)} \phi\left[a(u)x+b(u)y\right]
\ee
contains one arbitrary function $\phi$ of the argument $a(u)x+b(u)y$.
We shall assume this function to be non-vanishing so that $R_{ab}\neq
0$. We can now insert Eq.~\eqref{f} into the Poisson equation and
solve for $H(x,y,u)$.  We need two boundary conditions to determine
$H$ from the Poisson equation [Eq.~\eqref{poisson}] and another one to
determine the arbitrary function $\phi$ in Eq.~\eqref{f}.  Let us then
provide an example by assuming that $b(u)<0$ and $\phi$ remains
bounded. In this case, we must restrict the range of the coordinates
to the half-plane, $0\leq x <\infty$, $-\infty<y<\infty$. We impose a
boundary condition $H_0(u,y):=H(u,0,y)$ and appropriate fall-off
behavior for $|y|\to\infty$. We then obtain the particular solution
\ba
H(u,x,y)&=&\frac{1}{\pi}\int\limits_{-\infty}^{\infty}
\frac{xH_0(u,\eta)d\eta}{x^2+(y-\eta)^2} 
\nonumber \\
&-&
\frac{1}{4\pi} \int\limits_0^\infty\int\limits_{-\infty}^\infty
e^{-\xi/|b(u)|} \phi\left[a(u)\xi+b(u)\eta\right]
\nonumber \\
&\times&
\ln{\left[\frac{(x+\xi)^2+(y-\eta)^2}{(x-\xi)^2+(y-\eta)^2}\right]}\,d\xi 
d\eta   
\label{eq:95}
\ea 
where the double integral extends over the half-plane. 

%One can show straightfowardly that when $a(u) = a = {\rm{const.}}$,
%$b(u) = b = {\rm{const.}}$ and $H_0 = 2 h_0 \delta(u) \ln{y}$ the new
%solution becomes

The exponential behavior in Eq.~\eqref{f} is particularly interesting,
since it resembles the gravitational wave solutions found in
Refs.~\cite{Jackiw:2003pm,Alexander:2004wk,Alexander:2007zg,Alexander:2007:bgw}.
Moreover, as $x\to\pm\infty$ [depending on the sign of $b(u)$] the
source term in Eq.~\eqref{poisson} diverges, indicating a possible
instability. Since we were mainly concerned with the existence of
solutions we have not attempted to construct solutions for more
general $\theta$ than Eq.~\eqref{simplepptheta}.

%-----------------------------------------------------------------
\subsection{Losing a Killing Vector}\label{se:loser}

From the analysis so far, it is clear that stationary and axisymmetric
solutions in CS gravity do not seem to be capable of describing
physical spinning BHs. The far-field solution has guided us in the
direction of loss of axisymmetry, which in essence corresponds to
losing the $(\partial_{\phi})^a$ Killing vector. Analogously we could
conceive of losing stationarity instead of axisymmetric by dropping
the $(\partial_{t})^a$ Killing vector. The general idea is then that
by losing one Killing vector we gain new undetermined metric
components that could allow for a physical spinning BH solution in CS gravity. However, our
attempts have not revealed any interesting exact solution
corresponding to a spinning BH, so we confine ourselves to a couple of
general remarks.

Spinning BHs that break axisymmetry or stationarity would be radically
different from those considered in GR. On the one hand,
non-axisymmetric spinning objects would have an intrinsic precession
rate that would not allow the identification of an axis of rotation.
Such precession would possibly also lead to solutions with more than
two non-zero multipole moments, thus violating the no-hair theorem. On
the other hand, non-stationary spinning objects would unavoidably lead
to the emission of gravitational radiation,
even if the BH is isolated. These considerations could be flipped if
we take them as predictions of the theory, thus leading to new
possible tests of CS gravity.  Work along these lines is currently
underway. The results of \cite{Grumiller:2007} for the Pontryagin
constraint may be helpful here.

%-----------------------------------------------------------------
\subsection{Adding matter}\label{se:whatsthematter}

The inclusion of matter sources is of relevance in the present context
for several reasons. First, the Kerr BH has a distributional energy
momentum tensor \cite{Balasin:1994kf}, so we need not set the
stress-energy tensor strictly to zero to construct a Kerr-like
solution.  Second, in Ref.~\cite{alexander:2004:lfg} the
Pontryagin-term in the action arises from matter currents, so the
inclusion of the latter would actually be mandatory within that
framework.

Two conceptually different approaches are possible to the problem of finding exact
solutions of the modified field equations in the presence of matter.
These approaches essentially depend on whether we require the
energy-momentum tensor to be covariantly conserved, $\nabla_a
T^{ab}=0$, or not. If this
tensor is conserved, then the Pontryagin constraint must be satisfied
and the Kerr BH cannot be a solution. Basically, this route leads to
only a slight generalization of the discussion presented so far, with
solutions of class $\mathcal{P}$,
\be 
R_{ab}-\frac12 g_{ab}R=8\pi T_{ab}\,,\qquad C_{ab}
= 0\,, 
\ee
and solutions of class $\mathcal{CS} \; \backslash \; {\cal{P}}$ that
solve Eq.~\eqref{eq:eom}.  Relaxing covariant conservation of the
stress-energy tensor, we can promote the Kerr BH to a solution of the
modified field equations, provided that
\be 
R_{ab}-\frac12 g_{ab}R=8\pi T_{ab}^{\rm
  dist}\,,\qquad C_{ab} = 8\pi T_{ab}^{\rm ind}\,.
\label{eq:ind}
\ee
Here $T_{ab}^{\rm dist}=0$ except for the usual distributional
contributions for Kerr \cite{Balasin:1994kf}, while $T_{ab}^{\rm ind}$
provides the non-conserved matter flux. The induced matter
fluxes for the Kerr BH are given by 
\begin{align}
\label{eq:C1}
T^{\rm ind}_{tr} &= \frac{a m^2}{4 \pi \mu \Sigma^5 \Delta} \cos{\Theta} \left(r^2
  - a^2 \cos^2{\Theta} \right) 
\nonumber \\
& \left[ a^2 \cos^2{\Theta} \left(11 r^2
    - a^2 \right) + r^2 \left(3 r^2 - 9 a^2 \right) \right]\,,
\\
T^{\rm ind}_{t \Theta} &= - \frac{a m^2 r}{4 \pi \mu \Sigma^5} \sin{\Theta} \nonumber \\
& \left[ 3 r^4 - 12 r^2  a^2 \cos^2{\Theta} + a^4 \cos^4{\Theta} \right]\,,
\\
T^{\rm ind}_{\phi r} &= -\frac{a^2 m^2}{4 \pi \mu \Sigma^5 \Delta} \sin^2{\Theta}
\cos{\Theta} \left[\cos^4{\Theta} a^4 \left(a^2 - r^2\right) 
\right. 
\nonumber \\
&+ \left. 
  \cos^2{\Theta} r^2 a^2 \left(8 a^2 + 12 r^2 \right) - 9 r^4 a^2 - 3  
  r^6 \right]\,,
\\
T^{\rm ind}_{\phi\Theta} &= -a\sin^2{\Theta}\, T^{\rm ind}_{t \Theta}\,.
\label{eq:C4}
\end{align}
Of course, with such a method any GR solution can be promoted to a
solution of the modified field equations. 

The crucial issue here is whether or not the induced matter flux can
be regarded as physically acceptable. In order to shed light on this
issue, we analyzed if the induced stress-energy given by
Eqs.~\eqref{eq:C1}-\eqref{eq:C4} obeys the energy conditions of GR
\cite{Hawking:1973}. Because $T_{ab}^{\rm ind}$ is always traceless,
the strong and weak energy conditions are equivalent and reduce to the
statement that $T^{\rm ind}_{ab}\xi^a\xi^b\geq 0$ for any timelike
vector $\xi^a$. This, however, is not the case, as we can show by
considering for instance $\xi^t=\sqrt{2}$, $\xi^r=1$, which is
timelike for sufficiently large $r$: $\xi^a\xi^b g_{ab}=-1+6
m/r+{\cal{O}}(m/r)^2$.  The only relevant component of $T_{ab}^{\rm
  ind}$ is given by Eq.~\eqref{eq:C1}, but since $T^{\rm ind}_{tr}$ is
proportional to $\cos{\Theta}$, this quantity is negative in half of
the spacetime, and thus the weak energy condition is violated.  While
this might be tolerated close to the horizon, we stress that this
violation arises also in the asymptotic region. This violation is
somewhat attenuated by the fall-off behavior of $T_{ab}^{\rm ind}$,
where its components decay at least as $1/r^5$ and the scalar
invariant $T_{ab}^{\rm ind} T^{ab\;\rm ind}$ as $1/r^{12}$ as $r \to
\infty$. Thus, if ordinary matter is added then the induced exotic
fluxes might not be detectable after all for a far-field observer.

There is another approach capable of circumventing the Pontryagin
constraint that also relies on new matter sources. Namely, if the
field $\theta$ is considered a dynamical field, instead of an external
field, it is natural to study more general actions than
Eq.~\eqref{CSaction} with $S_{\rm mat}=0$, such as~\cite{Smith:2007jm}
\be 
S = \kappa \int d^4x \sqrt{-g} \left(R + \frac12 (\nabla\theta)^2 -
  V(\theta) - \frac{1}{\alpha} \theta \; \pont \right)\,.  \ee Then
the Pontryagin constraint \eqref{eq:constraint} is replaced by \be
\pont = - \alpha [\square \theta + V'(\theta)] \,,
\label{eq:newpont}
\ee
where $\alpha$ is a constant. This provides a natural generalization
of the model considered in our paper. However, it also introduces an
amount of arbitrariness, since $V$ is a free function and, in fact,
more general couplings between $\theta$ and curvature might be
considered.

We conclude that allowing GR solutions to be also CS solutions by
inducing a stress-energy tensor via Eq.~\eqref{eq:ind} can lead to
unphysical energy distributions. In particular, the Kerr solution
induces an energy momentum tensor given by
Eqs.~\eqref{eq:C1}-\eqref{eq:C4}, which violates all energy
conditions, even in the asymptotic region.  The alternative approach
described above lifts $\theta$ to a genuine dynamical field with a
kinetic term and possibly self-interactions, at the cost of
introducing an arbitrary potential.

%%%%%%%%%%%%%%%%%%%%%%%%%%%%%%%%%%%%%%%%%%%%%%%%%%%%%%%%%%%%%%%%%%%%
\section{Conclusions and discussion}
\label{conclusions}

No exact solution has yet been found that could possibly represent a
spinning BH in CS modified gravity. In particular, the Kerr solution
is found to be incompatible with the constraints imposed by the
modified field equations. Previously, only perturbative solutions of
CS gravity had been considered, which might represent the exterior of
a BH.  The first study was carried out by Alexander and
Yunes~\cite{Alexander:2007zg,Alexander:2007vt}, who performed a
weak-field parameterized post-Newtonian analysis to find a
non-axisymmetric Kerr-like solution.  This study was later extended by
Smith, et.~al.~\cite{Smith:2007jm} to non-point like sources, finding
that the Israel junction conditions are effectively modified by the
C-tensor. Another study was carried out by Konno,
et.~al.~\cite{Konno:2007ze}, but this analysis was restricted to a
limited class of perturbations that did not allow for the breakage of
stationarity or axisymmetry.  Within that restricted perturbative
framework, a Kerr-like solution was found only for non-canonical
choices of $\theta$, concluding that BHs cannot rotate in the modified
theory for canonical $\theta$.  This conclusion of Konno et.~al.~is at
odds with both the results of Alexander and Yunes and Smith et.~al.

In order to address these issues, in the current paper we attempted to
determine what replaces the Kerr solution in CS modified gravity. We
thus studied exact solutions of the modified theory, comprising
spherically symmetric, static-axisymmetric, and
stationary-axisymmetric vacuum configurations, as well as some
generalizations thereof. 

We began our analysis in Sec.~\ref{se:ABC} by considering the CS
action in detail and rederiving the equations of motion, together with
the resultant surface integral terms. We continued in
Sec.~\ref{se:pont} by rederiving the Pontryagin constraint from the
equations of motion and providing two alternative interpretations of
it. One of them [Eq.~\eqref{I2}] is a reality condition on a quadratic
curvature invariant of the Weyl tensor, while the other
[Eq.~\eqref{EB}] is a null condition on the contraction of the
electric and magnetic parts of the Weyl tensor.  Before considering
specific line elements, in Sec.~\ref{se:sol} we classified all
solutions into two groups: GR-type (class ${\cal{P}}$), which contains
solutions of both the vacuum Einstein equations and the modified field
equations; non-GR type (class ${\cal{CS}} \; \backslash \;
{\cal{P}}$), which contains solutions of CS gravity that are not
solutions of the vacuum Einstein equations (cf.~Fig.~\ref{sol-space}).

After these general considerations, we began a systematic study of
line elements, starting with general spherically symmetric metrics in
Sec.~\ref{persistence}.  This class of line elements
[Eq.~\eqref{spherical}] is particularly important since it contains
the Schwarzschild, Friedmann-Robertson-Walker and Reissner-Nordstr\"om
solutions. We showed that, for the canonical choice of the CS scalar
field [Eq.~\eqref{canon}] and more general choices
[Eq.~\eqref{possibletheta2}], the modified field equations decouple
and any possible solution is forced to be of class ${\cal{P}}$.

We continued in Sec.~\ref{static} with an analysis of static and
axisymmetric metrics [Eq.~\eqref{ds2}]. We showed that, for the
canonical choice of the CS scalar field and more general choices
[Eq.~\eqref{v-gen-static}], the modified equations decouple once more.
We also showed that any static and axisymmetric line element is forced
to become spatially conformally flat, provided the field equations
decouple.  Exploiting this result, we found three different solutions
[Eqs.~\eqref{Schw}, \eqref{eq:Per1} and \eqref{eq:Per2}], only one of
which was physically relevant, namely the Schwarzschild solution.  For
the most general CS scalar field, however, the field equations do not
decouple, but we have shown that fields with such generality do not
seem to allow for a solution to the field equations apart from trivial
ones.  Thus, we may conclude that CS gravity does not allow for static
and axisymmetric solutions, apart from flat space, the Schwarzschild
solution and two additional (unphysical) solutions, irrespective of
the choice of the CS scalar field.

Static line elements then gave way to the central point of this paper:
stationary and axisymmetric solutions of CS gravity, discussed in
Sec.~\ref{stationary}. As in the previous cases, we showed that, for
the canonical choice of the CS scalar field and slightly more general
choices [Eq.~\eqref{stationary-axi-theta}], the field equations again
decouple. In this case, however, the Pontryagin constraint does not
hold automatically and we used it to constrain the class of possible
metric functions, cf.~appendix \ref{app:A}. In essence, the decoupling
requires not only that solutions must obey the Einstein equations, but
also the fulfillment of additional constraints
(cf.~Ref.~\cite{nicoswebsite}), which leads to an overdetermined
system of PDEs. Therefore, we concluded that non-trivial stationary
and axisymmetric solutions do not seem to exist for canonical CS
fields.

When a completely generic CS scalar field is considered, the modified
field equations do not decouple and solutions are not easy to find,
even with the simplifications derived from the Pontryagin constraint.
However, generic CS fields increase the degrees of freedom of the
problem and thus might allow for stationary and axisymmetric
solutions. We proved this statement by providing an example in
Sec.~\ref{se:vS}, through a sub-class of stationary and axisymmetric
metrics [Eq.~\eqref{eq:vSle}], belonging to the van Stockum class. In
that case, we showed that the only possible CS field compatible with
the field equations excludes the canonical choice. Moreover, we found
both, non-flat solutions of class ${\cal{P}}$ [Eq.~\eqref{not-flat}]
as well as non-flat solutions of class ${\cal{CS}} \; \backslash \;
{\cal{P}}$ [Eq.~\eqref{not-GR}] To the best of our knowledge, this is
the first time an exact solution in CS modified gravity is constructed
that is not also a solution of GR.  One of these solutions
[Eq.~\eqref{not-GR}] represents mathematical BHs, in the sense that
although they exhibit a Killing horizon, they are not physically
relevant, because the Killing vector generating the 'axial' symmetry
is light-like and closed timelike curves arise. We concluded that it
is unlikely that stationary, axisymmetric solutions exist that
represent a spinning physical BH.

% For the
%convenience of the reader we present now a particular BH solution as
%an example, \be
%\label{eq:conclusion1}
%ds^2 = -\rho \left(1-\frac{2m}{\sqrt{\rho}}\right) dt^2 - 2 \rho dt d\phi + \frac{1}{\sqrt{\rho}} \left(d\rho^2 + dz^2\right)\,,
%\ee
%and
%\be
%\label{eq:conclusion2}
%\theta = \frac23 \sqrt{\rho}\,z\,.
%\ee
%It is a simple exercise to show that this configuration of line element and scalar field $\theta$ obeys the equations of motion \eqref{eom} and that the metric is not Ricci-flat. It is obvious by inspection that there is a Killing horizon at $\sqrt{\rho}=2m$. This vacuum solution of CS modified gravity is not a vacuum solution of GR, but would require an energy momentum tensor determined from
%\be
%\label{eq:conclusion3}
%R_{tt}=-\frac{m}{4\rho}\,.
%\ee

Finally, in Sec.~\ref{beyond} we considered the possibility of
constructing solutions beyond the set of stationary and axisymmetric
spacetimes. We began in Sec.~\ref{se:KCS} by considering CS scalar
fields whose velocity is a Killing vector of the spacetime and found
that the only interesting case arises if that vector is null.
Naturally, such considerations led to exact gravitational shock-wave
spacetimes [Eq.~\eqref{pp}].  Within this pp-wave scenario, in
Sec.~\ref{se:pp} we constructed a generating method through which any
pp-wave solution of GR can be lifted to a solution of CS modified
gravity with an appropriate choice of the CS scalar field. We also
built a solution of class ${\cal{CS}} \; \backslash \; {\cal{P}}$
[Eq.~\eqref{eq:95}] that is not a GR pp-wave solution but does satisfy
the CS modified field equations.
 
Through this detailed study of solutions in CS gravity we have
ascertained that at least two different limits of the Kerr BH are
solutions to the modified field equations, even though the Kerr BH is
not: the Schwarzschild limit and the Aichelburg-Sexl limit. The former
was already known to be a solution to the CS modified field equations,
but the latter, which includes ultrarelativistically boosted BHs, was
not.  The existence of these solutions concurs with the naive
expectations expressed at the end of Sec.~\ref{se:pont}. Moreover,
such expectations, together with the non-axisymmetric far-field
solution, point to the existence of a physical spinning BH solution in
CS gravity, provided spacetimes with only one Killing vector are
considered. We addressed this possibility briefly in
Sec.~\ref{se:loser}, but unfortunately such spacetimes are so general
that the modified field equations become prohibitively difficult, even
with the use of symbolic manipulation software.

Other possibilities of bypassing the Pontryagin constraint were
discussed in Sec.~\ref{se:whatsthematter}, since this constraint is in
essence responsible for the absence of interesting stationary and
axisymmetric solutions. First, we stated that obviously any (GR or
non-GR) solution formally can be lifted to a solution of the modified
field equations by allowing for arbitrary matter sources, and we
demonstrated the nature of these matter sources for the Kerr BH. We
found that the induced energy momentum tensor
[Eqs.~\eqref{eq:C1}-\eqref{eq:C4}] is exotic even in the asymptotic
region, but drops off rapidly with the radial coordinate. Second, we
mentioned the possibility that the CS scalar field $\theta$ might
acquire a kinetic term and self-interactions. In this case, the
Pontryagin constraint ceases to hold and is replaced by a dynamical
condition [Eq.~\eqref{eq:newpont}], relating the gravitational
instanton density to the (generalized) Klein-Gordon operator acting on
$\theta$.

We now conclude with a list of possible directions for future research
to which our current work may provide the basis.
\begin{itemize}
\item The number of physical degrees of freedom in CS modified gravity
  is not known yet. Various considerations appear to lead to
  contradictory expectations. On the one hand, the appearance of
  higher order derivatives in the action [Eq.~\eqref{eq:intro}]
  suggests that additional degrees of freedom should emerge. On the
  other hand, the appearance of an additional constraint
  [Eq.~\eqref{pont}] suggests that fewer degrees of freedom should
  arise. Actually, the linearization procedure suggests that these
  competing effects cancel each other and that there are two
  polarizations of gravitons, just like in GR,
  albeit with properties that differ from GR
  \cite{Jackiw:2003pm,Alexander:2007:bgw}.
\item The role of boundary terms induced in CS gravity for BH
  thermodynamics could be investigated more
  thoroughly~\cite{dan-nico}. Also here general considerations lead to
  contradictory expectations. On the one hand, new boundary terms that
  arise in CS gravity [Eq.~\eqref{eq:boundary}] differ qualitatively
  from those that arise in GR or in scalar tensor theories. Such
  boundary terms suggest modifications of BH thermodynamics, even for
  solutions whose line elements coincide with GR solutions, like the
  Schwarzschild spacetime. On the other hand, the Pontryagin
  constraint eliminates the CS contribution [Eq.~\eqref{eq:intro}] to
  the on-shell action, which suggests that BH thermodynamics is left
  unchanged, at least in the classical approximation.
\item Both previous issues can be addressed by a thorough Hamiltonian
  analysis, which is also of interest by itself and for exhibiting the
  canonical structure as well as the classical constraint algebra.
  Such a study would also be useful for numerical evolutions of BH
  binary spacetimes in CS gravity, which is currently being carried
  out~\cite{Wood:prep}.
\item While our discussion of stationary and axisymmetric solutions
  was quite comprehensive, a few issues are still open, which may be
  an interesting topic for mathematical relativists. For instance,
  while we were able to provide a proof that there are only three
  types of solutions for static and axisymmetric spacetimes (with the
  canonical choice for the CS scalar field), we could only provide
  good evidence, but no mathematical proof, that no further solutions
  exist for spacetimes that are stationary and axisymmetric.
\item Combining the evidence found in this paper with the far field
  solutions found previously, we concluded that spinning BHs should
  break either stationarity or axisymmetry (or both) in CS modified
  gravity. Perturbations away from axisymmetry were neglected
  in~\cite{Konno:2007ze}, although non-axisymmetric solutions can
  still represent spinning BHs, albeit with an inherent precession
  induced by the CS modification.  Therefore, future work could focus
  on finding exact spacetimes with a smaller amount of symmetries.
\item A manageable implementation of the Pontryagin constraint could
  be useful in many CS gravity applications. The brute force methods
  that led us to the formulas in appendix \ref{app:A} will render any
  generalization unintelligible. The considerations presented in
  Ref.~\cite{Grumiller:2007} provide such an implementation, but it
  has not been exploited so far in the construction of explicit
  solutions.
\item Far-field solutions of CS gravity that break stationarity could
  also be studied. These solutions could then be used as tests of the
  modified theory, through comparisons with gravitational-wave and
  astrophysical observations~\cite{Yunes:prep}.
\item Perhaps it is feasible to apply the method of matched
  asymptotic expansion for caged BHs
  \cite{Gorbonos:2004uc,Gorbonos:2005px} to the construction of
  spinning BH solutions in the present context. To this end, one would
  need an asymptotic expansion and a near horizon expansion of that
  BH. The former exists already, so it remains to construct the latter
  and perform the asymptotic matching.
\item Finally, it is worthwhile to consider not just vacuum solutions,
  but also solutions with matter sources, as outlined briefly in
  Sec.~\ref{se:whatsthematter}.
\end{itemize}
Certainly the range of issues that can be addressed has been extended
in a non-negligible way. Only through a better understanding of the
consequences and predictions of CS gravity will we be able to
determine the viability of the modified theory.

%%%%%%%%%%%%%%%%%%%%%%%%%%%%%%%%%%%%%%%%%%%%%%%%%%%%%%%%%%%%%%%%%%%%
\acknowledgments

We are grateful to Stephon Alexander and Roman Jackiw for encouraging
us to study this problem in the first place and for enlightening
discussions.  We would also like to thank Abhay Ashtekar, Henriette
Elvang, Alexander Hariton, Scott Hughes, Ralf Lehnert, Ben Owen,
Richard O'Shaughnessy, Carlos Sopuerta, Max Tegmark and Richard
Woodard for discussions and comments.  Most of our calculations used
the computer algebra systems MAPLE v.11 in combination with the
GRTensorII package~\cite{grtensor}.

DG is supported in part by funds provided by the U.S. Department of
Energy (DoE) under the cooperative research agreement
DEFG02-05ER41360. DG has been supported by the project MC-OIF 021421
of the European Commission under the Sixth EU Framework Programme for
Research and Technological Development (FP6).  

NY acknowledges the support of the Center for Gravitational Wave
Physics funded by the National Science Foundation under Cooperative
Agreement PHY-01-14375 and support from NSF grant PHY-05-55-628.

%%%%%%%%%%%%%%%%%%%%%%%%%%%%%%%%%%%%%%%%%%%%%%%%%%%%%%%%%%%%%%%%%%%%%
\begin{appendix}

%%%%%%%%%%%%%%%%%%%%%%%%%%%%%%%%%%%%%%%%%%%%%%%%%%%%%%%%%%%%%%%%%%%%%
\section{Proof of $\pont = {^\ast\!}C\,C$}
\label{app:new}

The equality
\be
\pont = {^\ast\!}C\,C
\label{eq:appnew1}
\ee
relates the Pontryagin term expressed as in Eq.~\eqref{pontryagindef} to the Weyl tensor
\be
C^{ab}{}_{cd}  :=  R^{ab}{}_{cd} - 2 \delta^{[a}_{[c} R^{b]}_{d]} + \frac13 \delta^a_{[c} \delta^b_{d]} R 
\label{eq:appnew2}
\ee
and its dual
\be 
{^\ast}C^a{}_b{}^{cd}:=\frac12
\epsilon^{cdef}C^a{}_{bef}\,.
\label{Cdual}
\ee
Equation \eqref{eq:appnew1} is quite simple to prove, but not entirely obvious. Indeed, we were
not able to find it in any of the standard textbooks, review articles
or papers on CS modified gravity. Therefore, we provide here a proof
by straightforward calculation. 

\begin{proof}
  Let us begin by inserting the definitions \eqref{pontryagindef}, \eqref{eq:appnew2} and \eqref{Cdual} into \eqref{eq:appnew1},
\be
\pont={^\ast\!}R^a{}_b{}^{cd} R^b{}_{acd}={^\ast\!}C^a{}_b{}^{cd} C^b{}_{acd} + \Delta\,.
\label{eq:appnew3}
\ee
where $\Delta$ is precisely the violation of Eq.~(\ref{eq:appnew1}).
Thus, if we can show that $\Delta$ vanishes in Eq.~\eqref{eq:appnew3}
we have proven Eq.~\eqref{eq:appnew1}. The quantity $\Delta$ contains
eight terms. Four of them are linear in the Weyl tensor. Two of these
terms are proportional to $C_{cdef}$ and two are proportional to
$C_{cdeb}$.  Since
\be
\epsilon^{cdef}C_{cdef} = \epsilon^{cdef}C_{cdeb} = 0,
\ee
these terms vanish. Each of the remaining four terms contains at least
two Kronecker $\delta$. These terms always lead to a contraction of
the Levi-Civita tensor, e.g.~of the form $\epsilon^{cd}{}_c{}^f=0$.
Therefore, also these four terms vanish and establish
\be
\Delta = 0 \,. %\,,
\ee
\end{proof}
%which concludes our proof.

\break

\begin{widetext}
\section{Pontryagin constraint}\label{app:A}

For the line element Eq.~\eqref{LPW} the Pontryagin constraint
Eq.~\eqref{eq:constraint} is given by ($\bar{w}:=e^{-2U}w$)
\be
0= A_0 \bar{w} + A_1 \bar{w}^3 + A_2 \bar{w}_{,\rho} + A_3 \bar{w}^2\bar{w}_{,\rho} + A_4 \bar{w}_{,\rho}^2 + A_5 \bar{w}_{,z} + A_6 \bar{w}^2\bar{w}_{,z} + A_7\bar{w}^2_{,z} + A_{8} \bar{w}_{,\rho}\bar{w}_{,z} + A_{9} \bar{w}_{,\rho z} + A_{10} (\bar{w}_{,\rho\rho} - \bar{w}_{,zz})
\ee
with 
\begin{align}
%
% ordered for \bar{w}: terms with no derivatives
%
A_0 &= -\rho^2\,A_1 %8\,\rho^3\big(U_{,\rho}U_{,z}(U_{,\rho \rho} - U_{,z z}) - U_{,\rho z} (U_{,\rho}^{2} - U_{,z}^{2}) + U_{,\rho}^{3}k_{,z}  - U_{,z}^{3}k_{,\rho} - U_{,\rho}^{2}U_{,z}k_{,\rho} + U_{,z}^{2}U_{,\rho}k_{,z} \big) -8\,\rho^2\,U_{,z} \big(U_{,\rho}^2 + U_{,z}^2 \big) 
%\nonumber \\
%
%& \quad 
+2\,\rho^2 \big(2U_{,\rho z}k_{,\rho} - (U_{,\rho \rho}-U_{,z z}) k_{,z}  - 2 U_{,\rho \rho} U_{,z}  + 2U_{,\rho z} U_{,\rho} + 4U_{,z} (U_{,\rho}^2 + U_{,z}^2)+ U_{,z}(k_{,\rho \rho}+k_{,z z})   \nonumber \\ 
& \; + 8U_{,\rho}U_{,z}k_{,\rho} - 6U_{,\rho}^{2}k_{,z} + 2 U_{,z}^{2}k_{,z} - 2U_{,z}(k_{,\rho}^{2}  + k_{,z}^{2})  \big) 
+2\,\rho\big( U_{,\rho}k_{,z} - U_{,z}k_{,\rho} + 2\,U_{,z}U_{,\rho} \big)
\\
A_1 &= 8\rho\,\big(U_{,\rho}U_{,z}(U_{zz}-U_{\rho\rho}) + U_{,\rho z} (U_{,\rho}^{2} - U_{,z}^{2}) - U_{,\rho}^{3}k_{,z} + U_{,z}^{3}k_{,\rho} + U_{,\rho}U_{,z}(U_{,\rho}k_{,\rho}-U_{,z}k_{,z}) \big)
+8\,U_{,z} \big(U_{,\rho}^2 + U_{,z}^2 \big) 
%
% only \rho derivatives
%
\\
%\end{align}
%\begin{align}
A_2 &= 2\,{\rho}^{3}\big( 2U_{,z z}U_{,z} + 2U_{,\rho z}U_{,\rho} + 4U_{,\rho}^{2}U_{,z} + 4U_{,z}^{3} + (U_{,\rho \rho} - U_{,z z})k_{,z} - U_{,z} (k_{,\rho \rho} + k_{,z z}) -  2U_{,\rho z}k_{,\rho} - 4U_{,\rho}U_{,z}k_{,\rho} \nonumber \\ 
&\; + 2U_{,z}k_{,\rho}^{2} + 2U_{,z}k_{,z}^{2} - 4U_{,z}^{2}k_{,z}\big) 
-2\,{\rho}^{2} \big(U_{,\rho z} + 4U_{,z}U_{,\rho} - 2U_{,\rho} k_{,z}\big)
-\rho\,k_{,z}  \\
A_3 &= 4\,\rho\big(
3 U_{,\rho}^{2}k_{,z}
+U_{,z}^{2}k_{,z}
-2U_{,\rho}U_{,z}k_{,\rho} 
-4U_{,\rho}^{2}U_{,z}
+(U_{,\rho \rho}-U_{,z z})U_{,z}
-2U_{,\rho}U_{,\rho z} 
-4U_{,z}^{3}\big) - 8 U_{,\rho}U_{,z} \\
A_4 &= 2\bar{w} \big( \rho\, U_{,z} k_{,\rho} - 3\,\rho\,U_{,\rho}k_{,z} + U_{,z} + 6\,\rho\,U_{,\rho}U_{,z} + \rho\,U_{,\rho z}\big) 
- \bar{w}_{,\rho} \,\rho \big( 2U_{,z} - k_{,z} \big)
\\
%\end{align}
%\begin{align}
%
% only z-derivatives 
%
A_5 &= 2\,{\rho}^{3}\big(-2U_{,\rho \rho}U_{,\rho}-2U_{,\rho z}U_{,z}
-4U_{,z}^{2}U_{,\rho}-4U_{,\rho}^{3} + (U_{,\rho \rho}-U_{,z z}) k_{,\rho}
+U_{,\rho}(k_{,z z}+k_{,\rho \rho}) + 2U_{,\rho z}k_{,z} +4U_{,z}U_{,\rho}k_{,z}  \nonumber \\ & \;
-2U_{,\rho} (k_{,\rho}^{2}+k_{,z}^{2})
+4U_{,\rho}^{2}k_{,\rho}
 \big) 
+2\,{\rho}^{2}\big(U_{,\rho \rho}+4U_{,\rho}^{2}-2U_{,\rho}k_{,\rho}\big) %\nonumber \\ &\;
+{\rho}^{2}\big(2k_{,\rho}^{2}+2k_{,z}^{2} - k_{,\rho \rho} - k_{,z z}\big)
-\rho\,(2U_{,\rho}-k_{,\rho})
\\
A_6 &=4\,\rho\,U_{,\rho}(U_{,\rho \rho}-U_{,z z})
+8\,\rho\,U_{,z}U_{,\rho z} 
-4\,U_{,\rho}^{2}-12\,U_{,z}^{2}
+4\,\rho\big(4U_{,\rho}(U_{,\rho}^2+U_{,z}^2)-U_{,\rho}^{2}k_{,\rho}+2U_{,z}U_{,\rho}k_{,z}-3U_{,z}^{2}k_{,\rho}\big)  
\\
A_7 &= \bar{w}\big(6\,U_{,z}-2\,\rho\,U_{,\rho z}-12\,\rho\,U_{,z}U_{,\rho} 
-2\,\rho\,U_{,\rho}k_{,z}  
+6\,\rho\,U_{,z}k_{,\rho} \big)
+\bar{w}_{,z} \,\rho \big(2U_{,\rho} - k_{,\rho}\big) - \bar{w}_{,z}
\\
%\end{align}
%\begin{align}
%
% both first derivatives
%
A_8 &= 2\bar{w}\,\rho\big( U_{,z z} - U_{,\rho \rho} + 6U_{,z}^{2} - 6U_{,\rho}^{2} - 2U_{,z}k_{,z} + 2U_{,\rho}k_{,\rho}\big)  + 4\bar{w} U_{,\rho} \nonumber \\ &\;
+\bar{w}_{,\rho}\,\rho\big(2U_{,\rho} - k_{,\rho}\big) - \bar{w}_{,\rho}
-\bar{w}_{,z}\,\rho\big(2U_{,z} - k_{,z}\big)
\\
%
% mixed second derivatives
%
A_9 &= 2\,{\rho}^{3}\big(U_{,z z}  - U_{,\rho \rho} + 2U_{,z}^{2} - 2U_{,\rho}^{2} + 2U_{,\rho}k_{,\rho} - 2U_{,z}k_{,z}\big) - 2\,{\rho}^2\, k_{,\rho} 
-4\bar{w}^{2}\,\rho\,\big(U_{,\rho}^{2} - U_{,z}^{2}\big) \nonumber \\
& \; +4\bar{w}\,\rho\big(\bar{w}_{,\rho}U_{,\rho}-\bar{w}_{,z}U_{,z}\big)
-(\bar{w}_{,\rho}^{2}-\bar{w}_{,z}^{2})\,\rho  
\\
%
% second \rho derivatives
% 
A_{10} &= 2\,{\rho}^{3}\big(U_{,\rho z} + 2 U_{,\rho} U_{,z} - U_{,\rho}k_{,z} - U_{,z}k_{,\rho}    \big)
+{\rho}^{2}k_{,z}  
+4\bar{w}^{2}\,\rho\,U_{,\rho}U_{,z}
-2\bar{w}\bar{w}_{,\rho} \,\rho\,U_{,z} 
-2\bar{w}\bar{w}_{,z}\,\rho\,U_{,\rho}
+\bar{w}_{,\rho}\bar{w}_{,z}\,\rho
%\\
%
% second z derivatives: turns out to be minus the second \rho derivatives!
%
%A_{11} &= 2\,{\rho}^{3}\left(U_{,\rho}k_{,z} + U_{,z} k_{,\rho}-U_{,\rho z} -2 U_{,\rho} U_{,z}\right)
%-{\rho}^{2}k_{,z} 
%-4\bar{w}^{2}\,\rho\,U_{,\rho} U_{,z}
%+2\bar{w} \bar{w}_{,\rho}  \,\rho\,U_{,z}
%+2\bar{w} \bar{w}_{,z} \,\rho\,U_{,\rho}
%-\bar{w}_{,\rho}\bar{w}_{,z} \, \rho\,.
\end{align}

\end{widetext}

\end{appendix}

%%%%%%%%%%%%%%%%%%%%%%%%%%%%%%%%%%%%%%%%%%%%%%%%%%%%%%%%%%%%%%%%%%%%%
\bibliographystyle{apsrev}
\bibliography{review}

\end{document}